\documentclass[%
 reprint,
 nofootinbib,
 amsmath,amssymb,
 aps,
  prfluids,
  twocolumn,
]{revtex4-2}

\usepackage{graphicx}
\usepackage{dcolumn,booktabs}
\usepackage{bm}

\usepackage{mathtools}
\usepackage{graphicx}
\usepackage{dcolumn}
\usepackage{bm,color,stmaryrd}
\usepackage{hyperref}
\usepackage{upgreek}
\usepackage{siunitx}

\DeclareSIUnit\dyne{dyn}

\usepackage[dvipsnames]{xcolor}

\def\b{\bm}

\def\e{\varepsilon}

\def\p{\b{p}}
\def\x{\b{x}}

\def\D{\mathrm{D}}

\def\yhat{\hat{\b{y}}}
\def\zhat{\hat{\b{z}}}

\def\R{\mathrm{R}}
\def\Pr{\mathrm{S}}

\newcommand{\myeqref}[2]{(\hyperref[#1]{\ref*{#1}#2})}
\newcommand{\myref}[2]{\hyperref[#1]{\ref*{#1}(#2)}}
\newcommand{\myrefnb}[2]{\hyperref[#1]{\ref*{#1}#2}}

\def\XXint#1#2#3{{\setbox0=\hbox{$#1{#2#3}{\int}$}
     \vcenter{\hbox{$#2#3$}}\kern-.5\wd0}}

\definecolor{editcol}{RGB}{200,20,130}

\graphicspath{{figures/}}

\begin{document}

\preprint{APS/123-QED}

\title{Autonomous route to the strange attractor: \\the many life stages of the chaotic bubblewheel}

\author{Michael Zhao}
\affiliation{Department of Mathematics, University of Wisconsin--Madison, Madison, WI USA}%
\author{Saverio E.\ Spagnolie}
\email{spagnolie@math.wisc.edu}
\affiliation{Department of Mathematics, University of Wisconsin--Madison, Madison, WI USA}
\affiliation{
 Department of Chemical \& Biological Engineering, University of Wisconsin--Madison, Madison, WI, USA
}%

\date{\today}

\begin{abstract}
A body floating atop a supersaturated fluid may accumulate bubbles along its underbelly, which can render the body rotationally unstable. But rotation can strip the surface of these bubbles when they make contact with the air above. To explore this coupling we perform experiments using cylinders floating on carbonated water. The cylinders exhibit an array of distinct dynamics, including constant rolling, periodic and aperiodic oscillation, chaos, and intermittent capsizing, which can be tuned by body size, mass distribution, and gas concentration. A continuum model for the bubble density dynamics, and Galerkin projection, tie the experiment to the generalized Lorenz system and its famed chaotic attractor. Even an extremely small center-of-mass offset can have a qualitative impact on the dynamics, and an offset of as little as a few percent can fully stabilize the system. The experiment represents a highly accessible physical realization of the Lorenz system, which, owing to continuous gas loss from the fluid to the air above, sweeps slowly without intervention across a classical bifurcation diagram in time.
\end{abstract}

\maketitle

\section{Introduction}
Supersaturated fluids contain more gas than they can hold at equilibrium, resulting in bubble formation and escape \cite{Rayleigh17,ep50,Scriven59}, with implications for systems as disparate as flowing magma \cite{Manga96,gs98,lnl04} and champagne \cite{bjj02,zx08,Liger12}. Bubbles on their own display a wide array of dazzling behaviors~\cite{Lohse18,lz25}. When free particles are introduced into such an environment, bubbles may nucleate and grow on their surfaces. These bubbles can confer additional lifting surface tractions, resulting in complex body dynamics. For bodies of a certain range of densities, such bubble formation can result in particle levitation against gravity, followed by bubble removal at the fluid-air interface, and body plummeting \cite{Planinsic04,Cordry98,ddl00,mglo12,mkcdknp14,pwvstnd23,scg24}. But bodies with large surface areas (relative to the maximum bubble size) tend not to `dance' vertically if they are restricted from rotating; the body continues to be held aloft by the bubbles which remain submerged beneath it~\cite{scg24}. 

In this paper, we examine the many life stages of a cylindrical body rotating at the surface of a supersaturated fluid, which we term the `chaotic bubblewheel'. The system is explored experimentally using cylindrical bodies placed at the surface of carbonated water, whose orientational dynamics are tracked using computer vision. We also propose a complementary mathematical model, which describes the surface bubble coverage as a continuum with its own dynamics. The full model is explored numerically, and Galerkin projection is used to derive a reduced-order system of equations. A system which, for symmetric bodies, reduces to the classical Lorenz system, along with its famed `strange attractor' or `chaotic butterfly' dynamics \cite{Lorenz63,Sparrow1982,Strogatz2015}. 

The Lorenz system was originally derived as a severe truncation of the Boussinesq approximation of convective flow, but its realizations include the dynamics of a heated hoop \cite{Welander67,Malkus72,gwr86,yym87}, a freely rotating sandwheel \cite{ttb13}, single mode lasers \cite{Haken75}, disc dynamos \cite{knobloch81,hsa96}, electric circuits \cite{co93}, Quincke rotors \cite{Quincke96,ll02,pll05}, and pilot-wave walking droplets \cite{thorb16,Durey20,vspsv21,pv23}. The `bubblewheel' is so named due to the connection with the chaotic waterwheel of Malkus and Howard \cite{Sparrow1982,kg92,ifso12,Strogatz2015}, wherein cups affixed to a wheel are filled in one location with water, and continuously lose water as they rotate around a central axis. Here, bubbles play an inverted role via their buoyancy, and tend to be removed on contact with the air above. The bubblewheel joins this family as a new physical realization, one in which the charging and discharging processes are spatially separated by a fluid-air interface.

In the present experiment, the environment's gas concentration, which drives the rate of bubble growth, steadily diminishes over time as gas escapes through the interface \cite{lsbrc15,scg24}. We will show that the exponential decay of the gas concentration carries the system autonomously, without external influence, through the classical bifurcation diagram for the Lorenz system. The result is a passage through numerous life stages, from a first bifurcation to steady rolling, to the onset of pre-chaotic rotation, arrival at a quasi-steady attractor, and to a late stage of sudden, intermittent capsizing events. The behaviors may be tuned by body size and gas concentration. Any asymmetry in the mass distribution, even if remarkably small, is found to play a surprisingly large role in determining the body stability and dynamics.

\begin{figure*}
    \includegraphics[width=0.8\linewidth]{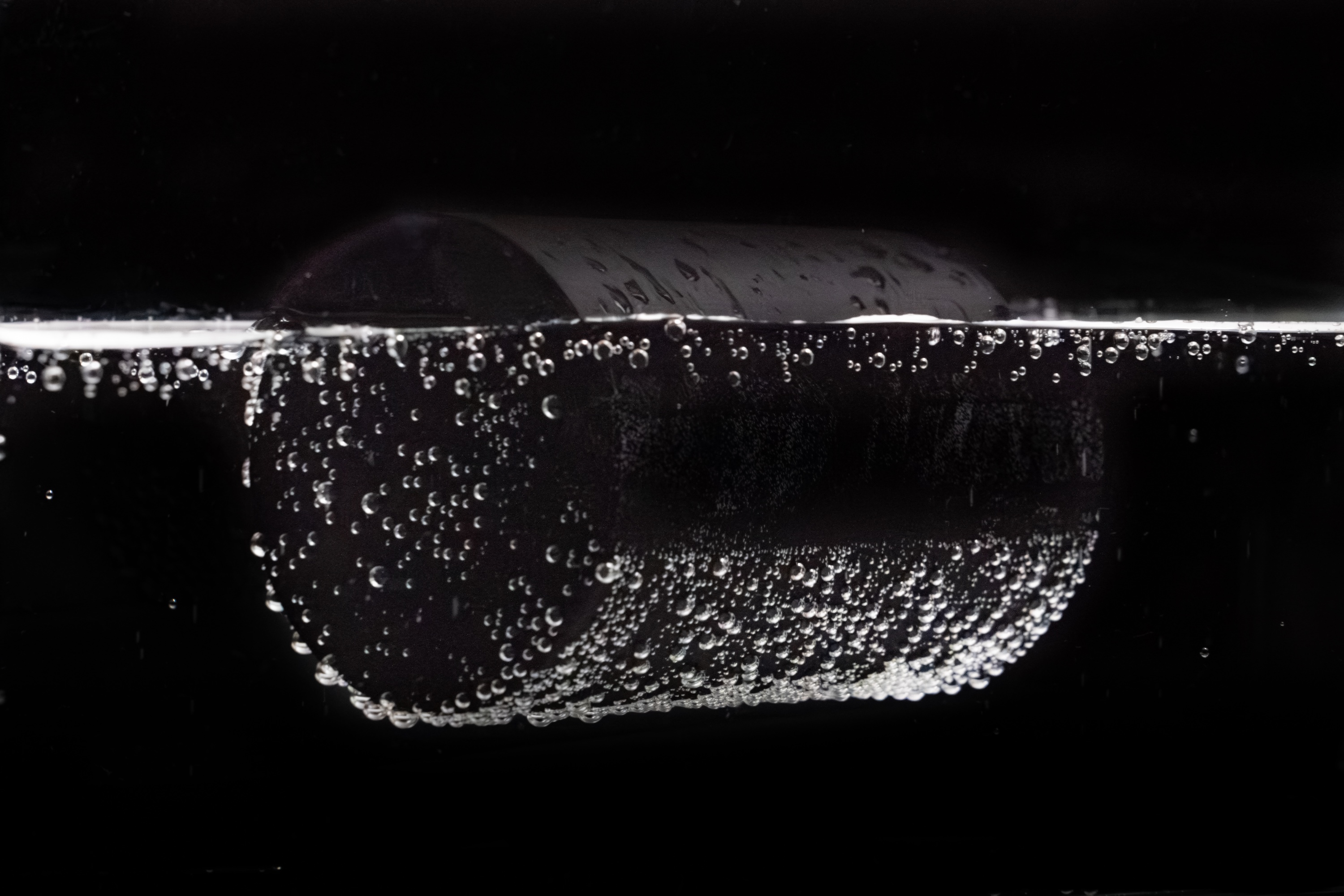}
    \caption{Bubbles form on the submerged surface of a positively buoyant cylinder of radius $a=2.5$cm and length $L=8.5$cm, which is floating atop carbonated water. The bubbles provide lift, but are removed from the surface when they meet the surrounding air. The body has just begun to rotate, and the bubble coverage is discontinuous across a sharp ridge, visible along the cylinder length. The surface above the ridge has just re-entered the fluid. The asymmetry in the bubble distribution on display generates a torque on the cylinder, which encourages continued body rotation in the same direction. Such instability to rotation leads to a rich space of dynamics. See Movie M0.}
    \label{fig: Figure1}
\end{figure*}

The paper is organized as follows. In \S II we describe the experimental methods and findings. In \S III we develop a mathematical model, based on a continuum approximation of surface bubble growth and the associated surface buoyancy effects. In \S IV, we derive by Galerkin projection a connection between the full model and the Lorenz system, and proceed to explore both the full and reduced-order models analytically and numerically, in \S V. We close with a discussion in \S VI.

\section{Experiments}\label{sec:Experiments}

\textbf{Setup.} Cylindrical bodies of radius $a$, length $L$, and mass $m$, were 3D printed using polylactic acid (PLA), a thermoplastic polyester. An infill density of 60\% was used, resulting in cylinders which were positively buoyant even in non-carbonated water. The mass of each cylinder, $m$, was measured using a scale accurate to $0.01$g, and we define the effective body density as $\rho_s = m/\pi a^2 L$. The center of mass offset, $d$, the distance between the center of mass and the geometric center, was measured by fitting the body's rotational dynamics in simple tap water (see \cite{SuppMat}). 

Experiments were conducted in a glass container filled with $550$ml of carbonated water, prepared using a SodaStream device at room temperature. The container was thoroughly cleaned before every trial. Shortly after the container was filled with carbonated water, a single cylinder was placed at its center, away from the walls, to reduce boundary effects. 

Video was recorded from $15$cm above the container, with a frame rate of $60$fps. The angular velocity was extracted from these overhead recordings using a computer vision pipeline, described in \cite{SuppMat}. Based on previous experiments, gas is expected to have escaped the carbonated water through the free surface roughly exponentially in time, on a timescale of approximately 25 minutes \cite{lsbrc15,scg24}. Experiments were thus performed for up to an hour, beyond which time most of the gas is believed to have escaped from the fluid. 

\begin{figure*}[t]
    \centering
    \includegraphics[width=\linewidth]{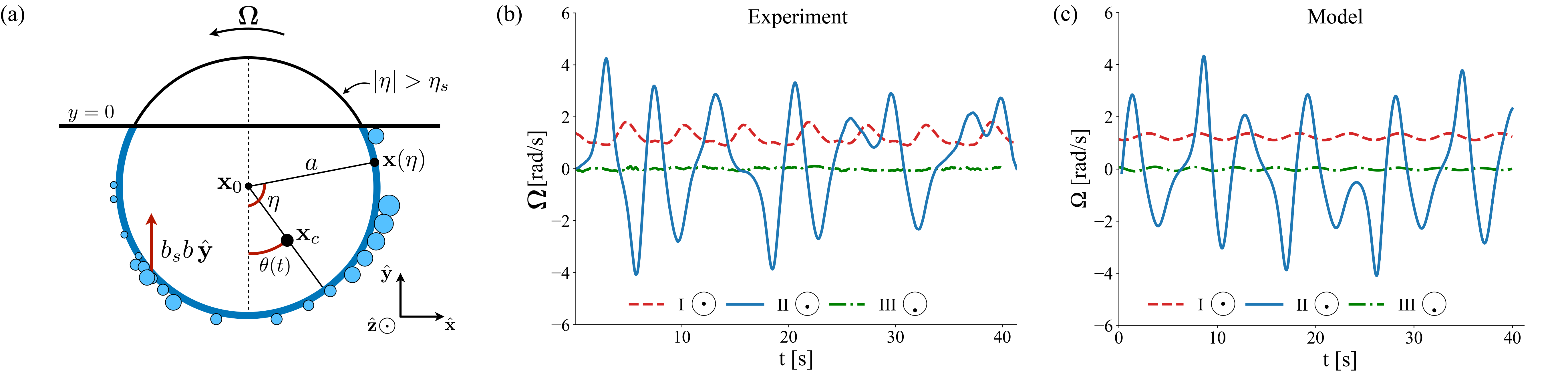}
    \caption{(a) Illustrated side-view of the experiment, and variable definitions. A cylindrical body of radius $a$ and centroid $\x_0$, is partially submerged in the fluid. The surface in the lab frame is parameterized by an angle $\eta$, with $|\eta|<\eta_s$ along the submerged arc, and $|\eta|>\eta_s$ along the exposed arc, with $\cos(\eta_s)=y_0/a$. The center of mass lies a distance $d$ away from the centroid at an angle $\eta=\theta(t)$, measured from the lowest point on the body. The proportion of surface bubbles relative to the maximum capacity is a field, $b(\x,t)\in[0,1]$, resulting in a traction $b_s b(\x,t)\yhat$ at position $\x$. (b) The experimentally measured rotational velocity, $\Omega$, as a function of time, for three cylinders of radius $a=2.5$cm but different mass distributions, in a bath of comparable gas concentration. Body I: A nearly uniform body ($d/a=0.05\%$) exhibited steady forward rolling with periodic undulations, driven by a gradient in the surface bubble density below (dashed line; see Movie M1). Body II: a slightly bottom-heavy cylinder ($d/a=0.5\%$) exhibited irregular, back-and-forth rotational dynamics (solid line; see Movie M2). Body III: yet larger mass asymmetry ($d/a=6\%$) suppressed the dynamics, and only small orientation fluctuations were observed throughout the experiment (dash-dotted line; see Movie M3). (c) Model predictions (from \S III) for the same three bodies in (b), show strong qualitative agreement.}
    \label{fig: Figure2}
\end{figure*}

\begin{figure*}
\includegraphics[width = 0.95\textwidth]{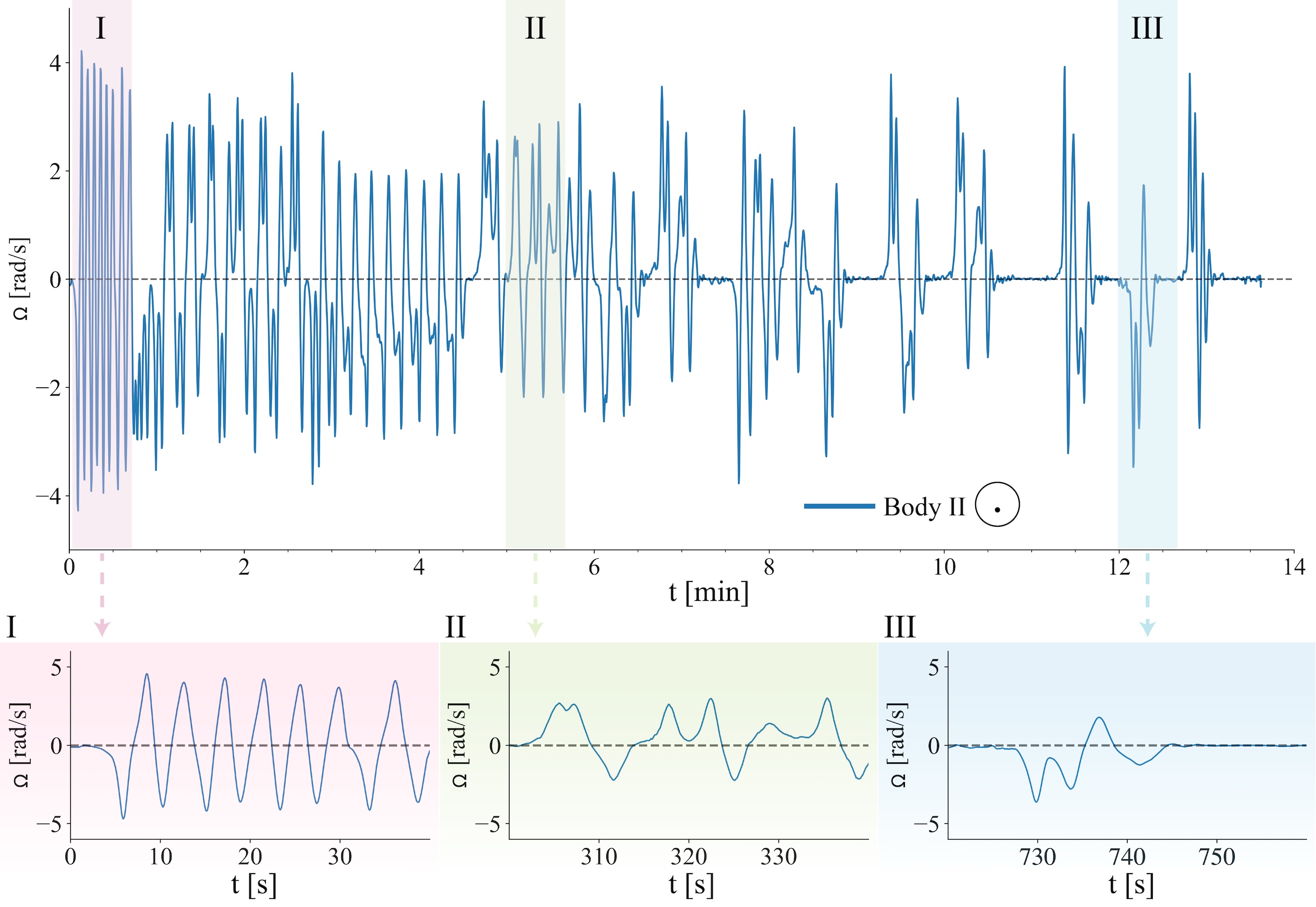}
\caption{The experimentally measured rotation rate of the slightly bottom-heavy Body II. The first 14 minutes of an hour-long experiment is shown. The system expresses numerous distinct dynamical regimes during a single experiment, owing to continuous gas escape to the air above. Early in the experiment (panel I), the dynamics are dominated by periodic oscillations due to the non-uniform mass distribution. At intermediate times (panel II), bubbles grow more slowly and chaotic behavior takes hold. At late times and very small bubble growth rates, the cylinder is nearly motionless, until it is disrupted by large, isolated capsizing events (panel III). See Movie M2.}
\label{fig: Figure3}
\end{figure*}

\textbf{Experimental results.} Upon a body's insertion onto the bath, bubbles were found to rapidly nucleate and grow on the submerged part of its surface. After a few seconds, body rotation was commonly observed. Bubbles which were subsequently carried out of the fluid tended to burst immediately. A representative example of the bubble coverage after such rolling has begun is shown in Fig.~\ref{fig: Figure1} (see also Movie M0). A sharp ridge along the length of the cylinder is visible below the waterline, above which bubbles have only just started to regrow. At the moment shown in the image, the asymmetric bubble distribution (illustrated in Fig.~\ref{fig: Figure2}a) contributes a torque which promotes continued acceleration in the same direction: a clear mechanism of rotational instability, and the fundamental driver of a wide variety of dynamical behaviors.

Bodies of different composition were used to explore a range of possible system dynamics. Figure \ref{fig: Figure2}b illustrates three (of the many) different behaviors observed in experiments, using three different bodies in baths of comparable gas concentration. All three bodies have radius $a=2.5$cm, and nearly the same mass, but differ in their internal mass distributions. For each, the body rotation rate, $\Omega$, is shown as a function of time, $t$, measured in seconds, starting a few seconds after inserting the body into the fluid.

Upon insertion, Body I, a large cylinder with roughly uniform mass distribution $(\rho_s,d/a) = (0.81,0.05\%)$, began rolling. It maintained an approximately constant angular velocity, with small but apparently periodic variations (the dashed line in Fig.~\ref{fig: Figure2}b; see Movie M1). Meanwhile, Body II, a very slightly `bottom-heavy' cylinder, with $(\rho_s,d/a)= (0.94, 0.5\%)$, showed qualitatively different behavior. The surface bubble coverage was similar to that for Body I, but this cylinder exhibited irregular, back-and-forth rotation (solid line; Movie M2). Finally, Body III, a cylinder with a more extreme mass asymmetry, with $(\rho_s,d/a)= (0.86, 6\%)$, remained essentially stationary throughout the experiment, with only very small fluctuations in its orientation over time (dash-dotted line; Movie M3). 

In each case, as the experiment proceeded, dissolved gas continuously escaped from the fluid through the fluid-air interface, and the bubble growth rate on the submerged surface steadily decreased, resulting in slow changes in the body dynamics over time. 

Figure \ref{fig: Figure3} shows the rotation rate of the slightly bottom-heavy Body II over the first 14 minutes of the experiment. Bubbles appeared to cover the submerged surface the instant it was placed into the bath. After the first few seconds, the body began to rotate back and forth, sweeping out almost the full range of body orientation angles, though not quite undergoing a complete rotation. The regularity of the pendulum-like rocking motion is observable in the rotation rate, shown in panel I. This behavior persisted for the first 45 seconds, after which it gave way to more highly erratic dynamics which continued for roughly 8 minutes. A brief selection from this period is enhanced in panel II. During this intermediate stage, the oscillations became irregular in both amplitude and period, and the direction of rotation changed unpredictably, suggesting chaotic dynamics. After the first 10 minutes of the experiment, the bubble growth rate having very noticeably reduced, the body entered another dynamical stage: minutes-long, nearly motionless stretches, interrupted by isolated `capsizing' events with large rotation rates, as shown in panel III.

A fourth body was also considered, one with a comparably small mass offset as Body I ($d/a=0.08\%$), but with a smaller radius ($a=1.5$cm). Like Body I, Body IV exhibited a long period (roughly 8 minutes) of steady rolling, notably with almost the same mean rotation rate ($\langle \Omega\rangle = 0.94$rad/s for Body I, while $\langle \Omega\rangle = 1.07$rad/s for Body IV, averaged between minutes 1 and 2 of the experiment; see \cite{SuppMat}).

More often than not during any such experiment, bubbles appeared to burst upon exiting the fluid bath, due to contact with the air. This was most plainly observed at small cylinder rotation rates. At large cylinder rotation rates, however, many bubbles did not burst during their brief excursion above the surface. Instead, they persisted in a thin liquid film, consistent with the Landau-Levich picture of viscous entrainment \cite{ll88}, and re-entered the bath on the opposite side. This phenomenon is clearly visible in Movie M4.

Taken together, these observations motivate a mathematical model that couples the surface bubble field to the body's rotational dynamics, with the aim of reproducing and understanding the various dynamical regimes during the lifetime of the experiment. 

\section{Continuum mathematical model} \label{sec:Modeling} 

To explore the coupled dynamics of bubble growth and cylinder rotation, we study the angular momentum balance governing the body rotation rate alongside a continuum approximation, following Ref.~\cite{scg24}, for the evolution of the surface bubble field. We begin by considering a cylindrical body of radius $a$, length $L$, mass $m$, and cross-sectional centroid position $\x_0(t)=(x_0(t),y_0(t))$ at time $t$, which is partially submerged in the fluid, as illustrated in Fig.~\ref{fig: Figure2}a. The fluid-air interface is assumed to be flat, and represented by $y=0$. We will confine our attention to cases in which $L/a\gg 1$, and treat each cross-section of the body (along the $z$ axis) as identical. 

\textbf{Kinematics.} The surface is parameterized by an angle $\eta\in(-\pi,\pi]$ in the lab frame, where $\x(\eta,t)=\x_0(t)+a(\sin \eta,-\cos \eta)$, and $\eta=0$ corresponds to the point on the body which is most deeply submerged. The intersection of the body and the fluid-air interface corresponds to angles $\eta=\pm \eta_s$, where $\cos(\eta_s)=y_0/a$; the submerged part of the body corresponds to angles $|\eta| \leq \eta_s$, and the exposed part to $|\eta| > \eta_s$. In general, $\x_0$ and thus $\eta_s$ are time dependent, though for this study they are assumed fixed. The center of mass, $\x_c(t)$, we write as $\x_c(t) = \x_0(t)+d\,\p(\theta(t))$, where $d:=|\x_c-\x_0|$ is the center of mass offset, and $\p(\theta)=\left(\sin\theta,-\cos \theta\right)$. If $d=0$, the angle $\theta(t)$ is retained and used to represent the change in body orientation relative to its initial state.

\textbf{Dynamics.} Angular momentum balance about the geometric center (centroid), $\x_0$, taken to be fixed, gives $d(J_0\,\Omega)/dt= T_0$, where $J_0$ is the moment of inertia about the centroid, $\Omega=d\theta/d t$ is the body's instantaneous rotation rate, and $T_0\zhat$ is the net torque acting on the cylinder. Since the cylinders were fabricated with nearly uniform density, with imperfections resulting in a very small mass distribution asymmetry, we take $J_0\approx ma^2/2$, with a relative error of $O(m(d/a)^2)$ as $d/a\to 0$. 

The net torque about the centroid has three contributions,
\begin{gather}
    T_0 = T_g + T_{v} + T_b,
\end{gather}
gravitational, viscous, and surface buoyancy effects, respectively. The gravitational torque is given by $T_g=-mgd\sin\theta$, where $g$ is gravitational acceleration. The viscous torque, $T_v$, is modeled as follows. A cylinder spinning in an infinite viscous fluid has an associated harmonic flow field, which exerts a torque on the cylinder which is linear in the rotation rate; and for a time-varying rotation rate, a viscous boundary layer contributes an additional torque which is linear in the rotational acceleration (a rotational `virtual mass' effect) \cite{Sedov65}. Combining these effects, we write
\begin{gather}
T_{v} = -4\pi \mu a^2 L \Sigma \dot{\theta}-\frac{\rho \pi a^4 L}{4}\ddot{\theta}.
\end{gather}
Here $\rho$ is the fluid density, and the coefficient $\Sigma$ is meant to capture the effects of viscous momentum diffusion through the fluid, and any free surface contributions to rolling resistance (e.g. surface tension). 

\begin{figure*}[t]
    \centering
    \includegraphics[width=\linewidth]{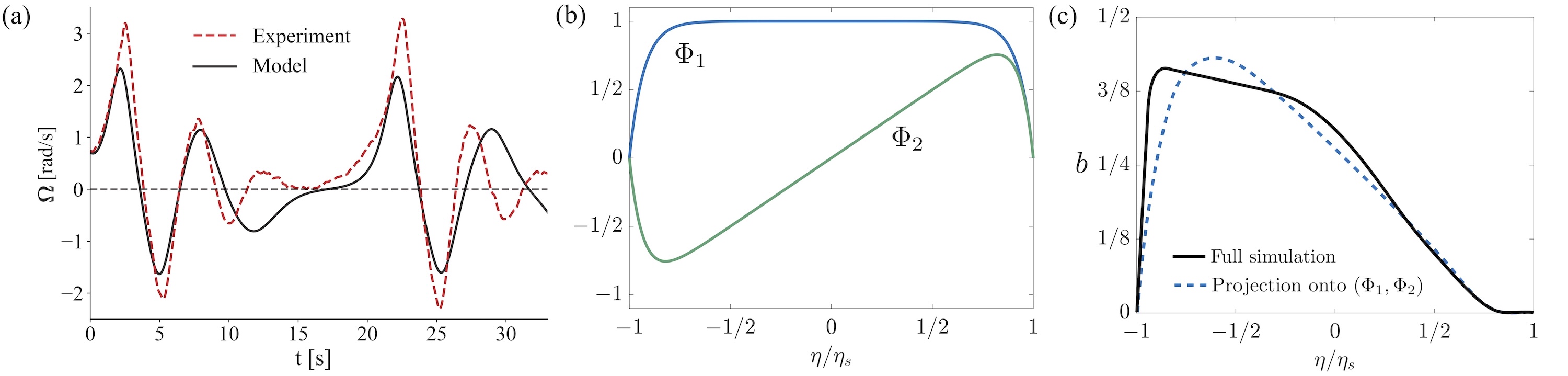}
    \caption{(a) Comparison between experimental measurements (dashed line) and model predictions (solid line) of the rotation rate as a function of time, using the slightly bottom-heavy Body II. The two show reasonably good agreement for roughly 25 seconds, capturing the amplitude and period of rolling and oscillations. But over time, the trajectories diverge, reflecting the system's sensitivity to small perturbations, and suggesting chaotic behavior. (b) The orthogonal basis functions, $\Phi_1$ and $\Phi_2$, for $\lambda/\varepsilon=50$, upon which we project the full continuum bubble field to derive a reduced-order model. (c) An example of the instantaneous $b$-field from the full simulations, and the best approximation from the space spanned by $(\Phi_1,\Phi_2)$ at that moment.}
    \label{fig: Figure4}
\end{figure*}

Finally, to model $T_b$, the torque due to the buoyancy of the attached bubbles, we first choose to treat the surface bubble field as a continuum. The surface has a maximum surface buoyancy traction with magnitude denoted by $b_s$. Bubble pinch-off and departure limits the surface traction from growing beyond these values \cite{Fritz35,ks14,scg24}. The surface bubble field is then described by the local fraction to the maximal capacity, $b(\eta,t)\,\in[0,1]$. The total associated body torque is found by integrating the local contributions,
\begin{gather}
  T_b= \int_{\partial S} a\sin\eta\, (b_s\, b)\,dS = a^2 L\, b_s \tilde{T}_b[b],
\end{gather}
where $\partial S$ denotes the body surface, $dS=L a\,d\eta$ is an infinitesimal surface area element. The functional
 \begin{gather}
\tilde{T}_b[b] := \int_{-\eta_s}^{\eta_s} \sin\eta \, b(\eta, t) \, d\eta
\end{gather}
is the dimensionless surface buoyancy torque, where the integration domain is only the submersed arc, where the bubbles are positively buoyant.

The field of bubbles was found in Ref.~\cite{scg24} to impose, on a suddenly immersed body, a force which grew linearly in time, before saturating to its maximal value. In the present experiments, upon exiting the fluid we observed that bubbles were rapidly, but not immediately, removed by contact with the air. We thus model the buoyancy field as
\begin{gather}
\displaystyle\frac{\D}{\D t}\left(b_s b(\eta,t)\right) =\Lambda(1-b)\, 1_{|\eta|<\eta_s}-\Lambda' b \,1_{|\eta|\geq \eta_s},
\end{gather}
where bubbles grow on the submerged arc $|\eta|<\eta_s$ and are removed on the exposed arc $|\eta|>\eta_s$ (the indicator function is denoted by $1$). Here, $\D/\D t = \partial_t+\Omega(t)\partial_\eta$ is a material derivative, and the bubble growth and removal rates, $\Lambda$ and $\Lambda'$, have units of stress per time. In Ref.~\cite{scg24}, $\Lambda$ was found to decay roughly exponentially in time for over an hour, $\Lambda(t)\approx \Lambda(0)\exp(-t/t_g)$, with $t_g\approx 25$min.

\textbf{Nondimensionalization.}\label{sec: nondimensionalization} One timescale in the system is given by $t_c:=m/8\pi\mu L=\rho_s a^2/8\mu$. The system is made dimensionless by scaling time upon $t_c$, and defining the dimensionless time $\tau = t/t_c$, resulting in:
\begin{align}
    \frac{d^2 \theta}{d\tau^2} &=-\delta \sin\theta - \sigma\frac{d\theta}{d\tau} + \beta \tilde{T}_b[b], \label{theta_dyn}\\
 b_\tau +\omega b_\eta &=\lambda(1-b)\,1_{|\eta|<\eta_s}-\lambda' b \,1_{|\eta|>\eta_s} \label{b_dyn} ,
\end{align}
where $b_\tau:=\partial_\tau b$ and $\omega(\tau):= t_c\Omega=d\theta/d\tau$, with five dimensionless groups governing the system dynamics:
\begin{align}
\delta &= \left(1 + \frac{\rho}{2 \rho_s}\right)^{-1}2 g a^3\left(\frac{\rho_s}{8\mu}\right)^2\left(\frac{d}{a}\right),\,\,\,\,\\
\sigma &= \left(1 + \frac{\rho}{2 \rho_s}\right)^{-1}\Sigma,\\
\beta &= \left(1 + \frac{\rho}{2 \rho_s}\right)^{-1}\frac{\rho_s a^2 b_s}{32\pi \mu^2},\\
(\lambda,\lambda') &= \frac{t_c}{b_s}(\Lambda,\Lambda') = \frac{\rho_s a^2}{8\mu b_s}(\Lambda,\Lambda').
\end{align}
Estimating from Ref~\cite{scg24} that $\Lambda\approx 4-50$ dyn cm$^{-2}$ s$^{-1}$ (at early times, decaying exponentially in time thereafter) and $b_s \approx 60$ dyn/cm$^2$ (assumed independent of time), for the bodies used in the experiments we find $t_c=20-75$s, $\lambda=1-60$, and $\beta=7\times 10^{3}-2\times 10^{4}$ (body-specific values are given in \cite{SuppMat}).

Because the gas escape is itself a dynamical process, a closed, autonomous system is formed by appending the equation $\lambda_\tau = -(t_c/t_g)\lambda$. For the present experiments, $t_c/t_g \approx 0.02-0.05$, so dynamics occur on a short timescale compared to gas loss. We therefore take $\lambda$ to be constant when investigating the dynamics (a quasi-steady approximation).

The remaining parameters for which we do not have immediate estimates are the scaled center-of-mass offset $\delta$, the rotational drag coefficient, $\Sigma$ (or $\sigma$), and the bubble removal rate along the exposed arc, $\lambda'$. The first two of these parameters are estimated by fitting to experiments using non-carbonated water, as described in \cite{SuppMat}. The values $d/a$ given in \S\ref{sec:Experiments} were based on this calibration, and we found $\Sigma = 23-28$ across all of the fitting procedures over a wide range of body rotation rates and behaviors, or $\sigma = 15-17$. Once determined, the bubble growth and removal rates, $\lambda$ and $\lambda'$, are found by fitting from trials using carbonated water. The parameters for each body used in the experiments are given in a table in \cite{SuppMat}.


Figure \ref{fig: Figure2}c shows the model predictions for the same three bodies used in experiments in Fig.~\ref{fig: Figure2}b. For this qualitative comparison we solved \eqref{theta_dyn}-\eqref{b_dyn} numerically in the Lagrangian (body) frame, taking as initial data $b(\eta,0)=0.5$ and $\theta(0)=0$, $\omega(0)$ chosen to suppress transient behavior, and a bubble growth rate, $\Lambda= 4$ dyn cm$^{-2}$ s$^{-1}$. The model predictions show excellent qualitative agreement with the experimental findings, and reasonable quantitative agreement. 

A more direct comparison between the model predictions and the experimental measurements, for the slightly bottom-heavy Body II, is shown in Fig.~\ref{fig: Figure4}a. Here the initial orientation angle and rotation rate from the experiment were matched at the beginning of the comparison, denoted by the time $\tau=0$, and the bubble growth rate over this window ($\lambda=1.0$) was found by fitting, as described in \cite{SuppMat}. In this closer inspection we again see a reasonably good quantitative agreement between model and experiment in the amplitude and period at early times. However, towards the end of the 30 second time window the trajectories begin to diverge. We interpret this divergence as reflecting the system's sensitivity to initial conditions, suggesting chaotic dynamics.

\section{Connection to the Lorenz system} \label{sec:Reduced order}

The spatial discontinuity of bubble growth and removal introduces a substantial analytical challenge. Nevertheless, a simplistic projection offers a better understanding of the effects of the various dimensionless groups. To this end, we approximate the bubble coverage ($b$-field) as
\begin{gather}
    b(\eta,\tau) \approx  p(\tau) \Phi_1(\eta)+q(\tau)\Phi_2(\eta).
\end{gather}
The basis functions $\Phi_1$ and $\Phi_2$ are chosen by introducing a small diffusivity $\e$ to the dynamics of $b$, so that the equilibrium profile (the normalized solution to $\lambda(1-b)+\e b_{\eta\eta}=0$, with $b(\pm \eta_s)=0$, having taken $\lambda' \to \infty$), is given by
\begin{gather}
    \Phi_1(\eta) =1-\frac{\cosh[(\lambda/\e)^{1/2}\eta]}{\cosh[(\lambda/\e)^{1/2}\eta_s]}.
\end{gather}
For a second basis function, orthogonal to the first under the standard inner product, we take $\Phi_2(\eta) = (\eta/\eta_s) \Phi_1(\eta)$. The functions $\Phi_1$ and $\Phi_2$ are shown in Fig.~\ref{fig: Figure4}b, and Fig.~\ref{fig: Figure4}c shows a comparison between the instantaneous b-field from a full simulation, and its best approximation in $L_2$ among functions spanned by $(\Phi_1,\Phi_2)$. Equations for the generalized Fourier coefficients $p(\tau)$ and $q(\tau)$ are found by Galerkin projection. The following system emerges in limit as $\e \to 0$:
\begin{align}
\omega_\tau &= -\delta \sin\theta -\sigma  \omega + 2 \beta  k \, q,\\
p_\tau&= \lambda  (1-p) -\frac{1}{2 \eta_s }q\, \omega,\\
q_\tau &= -\lambda q+\frac{3 }{2\eta_s}p\, \omega\\
\theta_\tau &= \omega,
\end{align}
where $k=\eta_s^{-1}\sin \eta_s-\cos \eta_s$. The following change of variables puts this system on a very familiar footing. Defining $s=\lambda \tau$ and writing
\begin{align}
p(\tau) &= 1 - Z(s)/R,\,\,\,\,
q(\tau) = \sqrt{3} Y(s)/R,\,\,\,\,\\
\omega(\tau) &=  \chi  \lambda X(s),\,\,\,\,\theta(\tau) = \chi\Theta(s),
\end{align}
where $\chi=2\eta_s/\sqrt{3}$, $\R=3\beta k/\lambda \sigma \eta_s$, and with $\dot{X}:=X_s$ we find that
\begin{align}
\dot{X} &=-\tilde{\delta} \sin\left(\chi\Theta\right) + \Pr(Y - X),\label{eq: Lorenz1}\\
\dot{Y} &= \R X - Y - XZ,\\
\dot{Z} &= XY - Z,\\
\dot{\Theta} &= X\label{eq: Lorenz4},
\end{align}
where $\Pr = \sigma/\lambda$, and $\tilde{\delta}=\delta/\chi \lambda^2$
\footnote{Note that $\tilde{\delta}\propto \lambda^{-2}$, so any mass asymmetry rapidly carries the system further away from the classical Lorenz system dynamics over time.}. When $\tilde{\delta}=0$ this is the celebrated Lorenz system. In the original context, $\Pr$ is the Prandtl number and $\R$ is the Rayleigh number (and here the `geometric factor' is one) \cite{Lorenz63}. When $\tilde{\delta}\neq 0$, the system is related to the Malkus-Robbins system or a 4D Lorenz system \cite{Robbins77,Moroz03,Moroz05}. The trigonometric term can be exchanged for a higher dimensional polynomial system by defining $F=\sin(\chi \Theta)$, $G=\cos(\chi \Theta)$, and then including $\dot{F}=\chi  G X$ and $\dot{G}=-\chi F X$.

Given this connection, we can lean on what is known about the Lorenz system to identify and predict behavioral transitions in the bubblewheel. 

In the Lorenz system, corresponding to the case of a uniform mass distribution, the fixed point $(X,Y,Z)=(0,0,0)$ remains stable for Rayleigh numbers $\R<1$, independent of $\Pr$. In the bubblewheel this corresponds to the state of no rotation, and a fully charged underbelly of bubbles, $b(\eta,t)\approx 1$. Counter-intuitively, perhaps, rapid bubble growth stabilizes the rotational dynamics. Any surface elements which are stripped of bubbles, resulting in an asymmetric $b$-field about $\eta=0$, (and giving a body torque), are immediately replenished with new bubbles, restoring symmetry.

At $\R=1$, however, the Lorenz system has a supercritical pitchfork bifurcation. For $\R > 1$, the motionless state above becomes unstable, and two new stable states emerge, corresponding here to steady rolling states. The fixed points of the Lorenz system are then $X=Y=\pm \sqrt{\R-1}$ and $Z=\R-1$. In the present experiment, this corresponds to rotation rates
\begin{gather}
\omega = \pm 2\sqrt{\frac{\beta  k}{\sigma }(\eta_s\lambda) -\frac{1}{3} (\eta_s \lambda )^2},
\end{gather}
where $\beta k/\sigma>\eta_s \lambda/3$ since $\R>1$, and $k \approx 1$ for a body seated deeply in the fluid. This bifurcation from motionlessness to rolling dynamics is predicted to occur once the bubble growth rate falls below the critical value
\begin{gather}
\lambda_{\text{roll}}=\frac{3\beta k}{\sigma \eta_s}.
\end{gather}

The predicted behavior for $\R>1$ depends on $\Pr$. If $\Pr<2$, the rolling states above are stable. If $\Pr>2$, however, a far richer story unfolds: as $\R$ is further increased, the dynamics begin to signal the onset of chaos. The steady rolling states lose stability through a sequence of transitions that includes transient chaos and phase-slipping. At a critical value, $\R_H$, the Lorenz system then undergoes a subcritical Hopf bifurcation, and beyond this point the dynamics are firmly chaotic, revealing the classical chaotic butterfly trajectory in the $(X,Y,Z)$ space. This critical value depends on $\Pr$, namely:
\begin{gather}
\R_H = \frac{\Pr(\Pr+4)}{\Pr - 2} = \frac{\sigma(\sigma+4\lambda)}{\lambda(\sigma-2\lambda)}
\end{gather}
(see \cite{Sparrow1982}). Since both $\Pr$ and $\R_H$ depend on the growth rate, $\lambda$, we can use this criterion to predict the bubble growth rate below which chaos is expected,
\begin{gather}
    \frac{3\beta k}{\lambda \sigma \eta_s} =\frac{\sigma(\sigma+4\lambda)}{\lambda(\sigma-2\lambda)},
\end{gather}
which is reached at a bubble growth rate
\begin{gather}
    \lambda_{\text{chaos}} = \frac{\sigma  \left(3\beta k-\eta_s\sigma^2\right)}{2 \left(3\beta k+2 \eta_s\sigma ^2\right)}.
\end{gather}
In the strongly forced regime  relevant to the experiments, $3k\beta\gg \eta_s \sigma^2$, we have $\lambda_{\text{chaos}}\approx \sigma/2$ (i.e.~when $\Pr=2$). 

\begin{figure*}
    \centering
    \includegraphics[width=0.98 \linewidth]{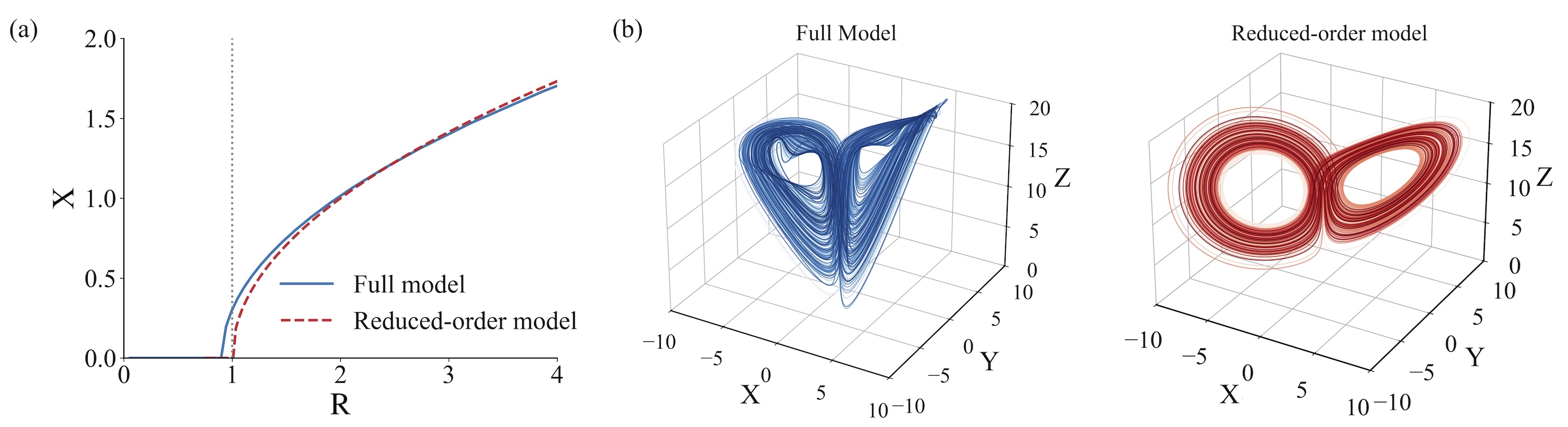}
    \caption{(a) Comparison of the steady, scaled body rotation rate ($X= (\chi \lambda)^{-1}\omega$) computed/predicted by the full/reduced-order models, as a function of the Rayleigh number $\R = 3\beta k / (\lambda \sigma \eta_s)$, with $\Pr/\R=14.06$ fixed. The full system exhibits a supercritical pitchfork bifurcation to steady rolling just below the critical value in the Lorenz system, $\R = 1$. (b) Trajectories in the Lorenz coordinates from simulating the full model (left) and the reduced-order model (right) after an initial, transient period, for fixed $\R=14.6$ and $\Pr=8$. The trajectory in the full model is qualitatively similar to the classical `strange attractor' of the Lorenz system; the chaotic dynamics of the bubblewheel are thus qualitatively captured by the low-dimensional projection.}
    \label{fig: Figure5}
\end{figure*}

Inserting the definitions of the dimensionless groups, the dimensional bubble growth rate below which the first pitchfork bifurcation to rolling behavior is predicted (for $\delta=0$, symmetric bodies) is
\begin{gather}
    \Lambda_{\text{roll}} = \frac{3k b_s^{2}}{4\pi\mu \eta_s \Sigma},
\end{gather}
which is notably independent of the body size, and the associated angular velocity when $\Lambda<\Lambda_{\text{roll}}$ is
\begin{gather}\label{eq: Omega_model}
    \Omega =\pm\frac{2 \eta_s\Lambda}{\sqrt{3}b_s} \sqrt{\frac{\Lambda_{\text{roll}}}{\Lambda }-1},
\end{gather}
which is also independent of the body size, since bubble forcing and rotational drag both scale with the surface area, consistent with the nearly identical rolling rates of the differently sized Bodies I and IV (see \cite{SuppMat}). 

Inserting the characteristic values provided in \S\ref{sec: nondimensionalization} gives rolling thresholds $\Lambda_{\text{roll}}\approx 10^{3}$ for both symmetric bodies, Body I and Body IV. This growth rate is orders of magnitude larger than those seen in the experiments. In addition, the Rayleigh number at the start of a trial is $\R\approx 25-350$, very far above the pitchfork (to rolling) threshold $\R=1$. Every body in the experiments is therefore predicted to be unstable to rolling from the moment of immersion, after a first, transient growth period lasting on the order of $b_s/\Lambda\approx 1-15$s. This is consistent with the observation that body rotation commenced within seconds of immersion into the bath. 

Since $\Pr=\sigma/\lambda$ can be small at early times in the experiment ($\Pr \approx 0.3$ initially for Body I, and $\Pr \approx 0.8$ for Body IV, for large but common initial bubble growth rates, \cite{SuppMat}), the steady rolling state is predicted to be stable for bodies of uniform mass distribution. And indeed, nearly steady rolling states were observed for up to $10$ minutes using both symmetric bodies, as evidenced in Fig.~\ref{fig: Figure2}b and Movie M1 (see also \cite{SuppMat}). 

Once $\Pr>2$, however, chaotic dynamics become possible. The bubble growth rate below which chaos is predicted in the strongly forced regime is
\begin{gather}
    \Lambda_{\text{chaos}} \approx\;\frac{4\mu\sigma b_s}{\rho_s a^2}.
\end{gather}
For the larger Body I we have $\Lambda_{\text{chaos}}\approx 8$, and for the smaller Body IV we have $\Lambda_{\text{chaos}}\approx 19$. For large but realistic initial bubble growth rates, $\Lambda > \Lambda_{\text{chaos}}$, Bodies I and IV are thus predicted to transition to chaotic behavior over the normal course of a single experiment, due to the exponential rate of gas loss from the environment. This too was commonly observed in experiments (see \cite{SuppMat} and Movie M1).

The Lorenz connection is specific to the case of bodies with uniform mass distribution. For asymmetric bodies, the critical Rayleigh number where the first bifurcation appears increases to a value $\R = 1+\chi \tilde{\delta}/(\Pr+1)$. Viewed differently, the fixed point with $(X,Y,Z,\Theta)=(0,0,0,0)$ can be stabilized by a critical mass asymmetry such that $\chi \tilde{\delta}=(\R-1)(\Pr+1)$ \cite{SuppMat}. This corresponds to a critical center-of-mass offset (taking $\R-1 \approx \R$) of
\begin{gather}
\frac{d}{a} \approx \frac{3 a k }{8 \pi \mu g \sigma \eta_s }\left(\Lambda+ \frac{\sigma b_s}{t_c}\right).
\end{gather}
For asymmetric Body III ($d/a\approx 6\%$), the critical growth rate below which the body is stabilized is $\Lambda \approx 70$ dyn cm$^{-2}$s$^{-1}$, far above the growth rates observed in the experiment. This explains the suppression of dynamics for Body III  throughout its lifetime, as seen in Fig.~\ref{fig: Figure2}b.

Numerical simulations were used to better understand the relationship between the full and reduced-order (generalized Lorenz) models, and to explore a wider range of the available parameter space. Figure \ref{fig: Figure5}a shows the (scaled) steady rolling angular velocity for symmetric bodies ($\delta=0$) as a function of the Rayleigh number, scanned by varying the bubble growth rate, $\lambda$. The parameters used for this study were $(\beta,\sigma,\eta_s)=(16,16,2.8)$. After a brief transient period, the dynamics in each case settled to the value shown (with symmetry broken by an small initial spin, $\omega(0)=0.01$). The full model shows a bifurcation just below $\R=1$, the critical value in the ideal, Lorenz setting. Figure \ref{fig: Figure5}b shows the system trajectory in time, after a transient period, for larger values of $\R$ and $\Pr$, namely $\R=14.6$ and $\Pr=8$, where theory predicts chaotic dynamics. The full model (left) shows a topologically similar attractor to the classical Lorenz butterfly (right). Both comparisons confirm that the Galerkin projection captures the essential features of the full system. Some discrepancy is expected due to the assumption of infinite bubble removal rate, $\lambda' \to \infty$, which is only valid for relatively small rotation rates, and due to the Galerkin projection.

\begin{figure*}
    \centering
    \includegraphics[width=0.98 \linewidth]{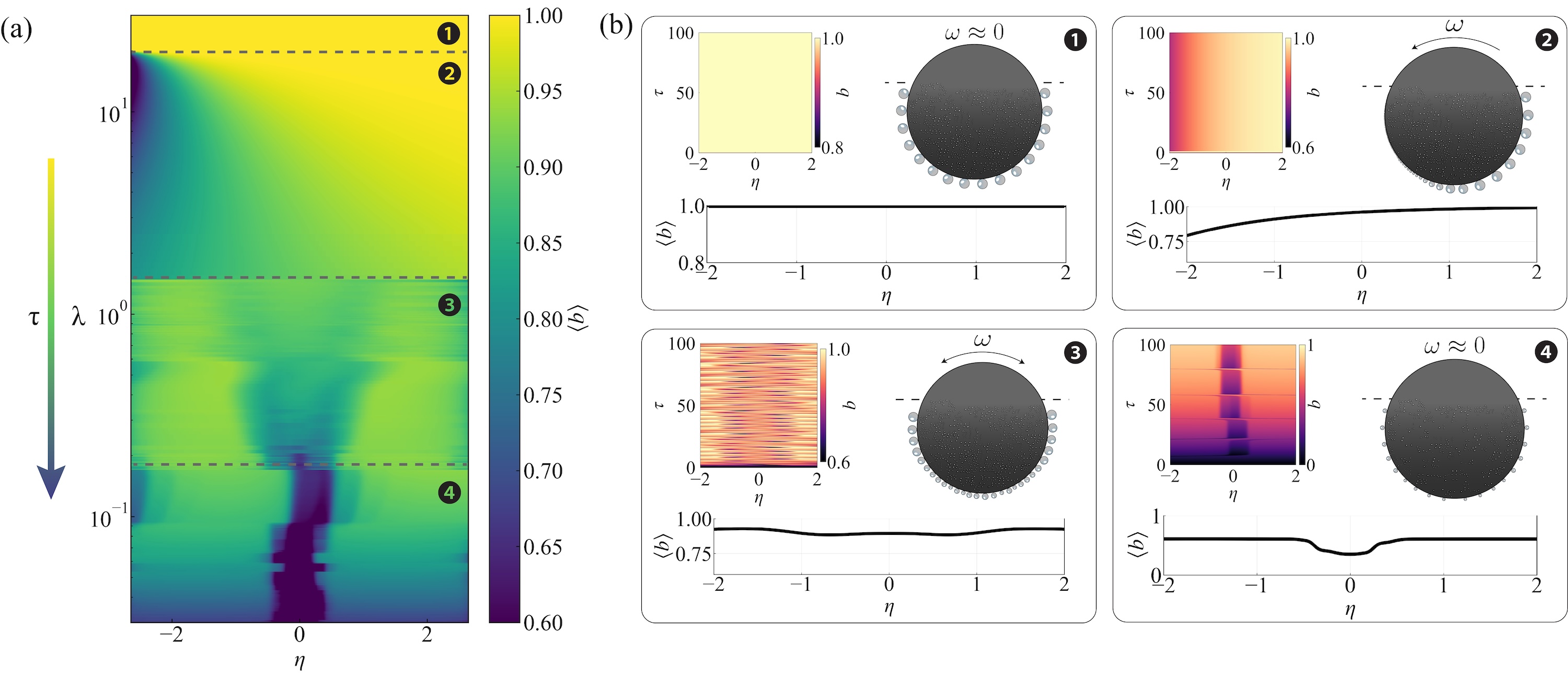}
    \caption{Symmetric bodies ($\delta=0$). (a) The time-averaged surface bubble coverage at each angle $\eta$, $\langle b \rangle(\eta)$, for parameters $(\beta, \sigma,\eta_s) = (160,10,2.8)$. As the bubble growth rate $\lambda$ decreases  (as time increases), the cylinder passes through four distinct life stages. The numbered stages are linked to the more detailed information provided in (b). (b) For each stage, the time evolution of $b(\eta, \tau)$ (above), the time-averaged profile $\langle b \rangle$ (below), and a schematic of the characteristic bubble distributions are shown. \textbf{Stage 1} ($\lambda=20$): rapid bubble growth saturates the surface uniformly, maintaining $\langle b \rangle \approx 1$ across all $\eta$; the cylinder remains stationary. \textbf{Stage 2} ($\lambda=8.32$): a persistent spatial gradient in the surface bubble distribution drives steady, unidirectional rolling. \textbf{Stage 3} ($\lambda=0.85$): the bubble field fluctuates irregularly in space and time, and the cylinder undergoes chaotic, reversing rotations with no preferred direction. \textbf{Stage 4} ($\lambda=0.02$): the cylinder is mostly stationary, until the immersed surface finally saturates with bubbles and a single large capsizing event is triggered.}
    \label{fig: Figure6}
\end{figure*}

\section{The life stages of the bubblewheel} \label{sec:Comparison of models} 

The model allows us to explore features of the system that are harder to measure experimentally, in particular, the structure of the (bubble) $b$-field and its relation to the rotational dynamics. We first focus on symmetric bodies, setting $\delta=0$. Figure \ref{fig: Figure6}a shows the pointwise, time-averaged surface coverage $\langle b\rangle(\eta) :=(1/100)\int_0^{100} b(\eta,\tau)\,d\tau$, as a function of the growth rate, $\lambda$. For this study, $(\beta, \sigma,\eta_s) = (160,10,2.8)$. Since $\lambda$ decreases (exponentially) as dissolved gas escapes, the vertical axis may also be viewed as a time axis. A single (numerical) experiment evolves from top to bottom, passing successively through four distinct stages, with more details provided in the four panels of Fig.~\ref{fig: Figure6}b. 

In stage 1 (early life; at large bubble growth rates), any patch of surface cleared by a small rotation is almost immediately refilled, so the submerged surface retains a nearly uniform bubble coverage at all times. The time-averaged field is therefore uniform and symmetric about $\eta=0$, produces no net torque, and the cylinder remains stationary. This corresponds to the stable fixed point at the origin of the Lorenz system. This hypothesized state was not observed in experiments, since the Rayleigh number $\R$ was large.

As gas escapes, the cylinder enters a second life stage, a steady rolling regime. With bubbles regrowing more slowly upon surface re-entry, a gradient in the bubble field develops, and the torque it produces turns the cylinder steadily in one direction. This is one of the two nontrivial Lorenz fixed points, and the initial perturbation sets which way the cylinder rolls. This behavior has also been observed in the related system of Quincke rotors, where the role of bubble growth is played by an induced interfacial electric charge \cite{Quincke96,Jones84,Turcu87,Vlahovska19,ds13}.

As the growth rate continues to fall, the cylinder enters a third stage, a chaotic regime, where bubbles regrow too slowly to maintain the steady gradient of stage 2. This behavior has been observed in Quincke rotors as well \cite{ll02,pll05}. After a significant rolling event, much of the surface is depleted of bubbles; the driving torque weakens, and rotation slows. As bubbles slowly regrow and redistribute, rotational dynamics begin anew, often in the opposite direction. Repeated many times, this interplay prevents the field from settling into a steady gradient, and replaces it with an irregular structure that drives the unsteady, reversing rotation. The time-averaged field in stage 3 is stronger near the waterline and weaker at the deepest point, owing to two contributions: bubbles which have grown large as they travel the full possible distance below the water during a single rotation, and bubbles which have survived a brief transit over the exposed arc.

\begin{figure*}
    \centering
    \includegraphics[width=0.98 \linewidth]{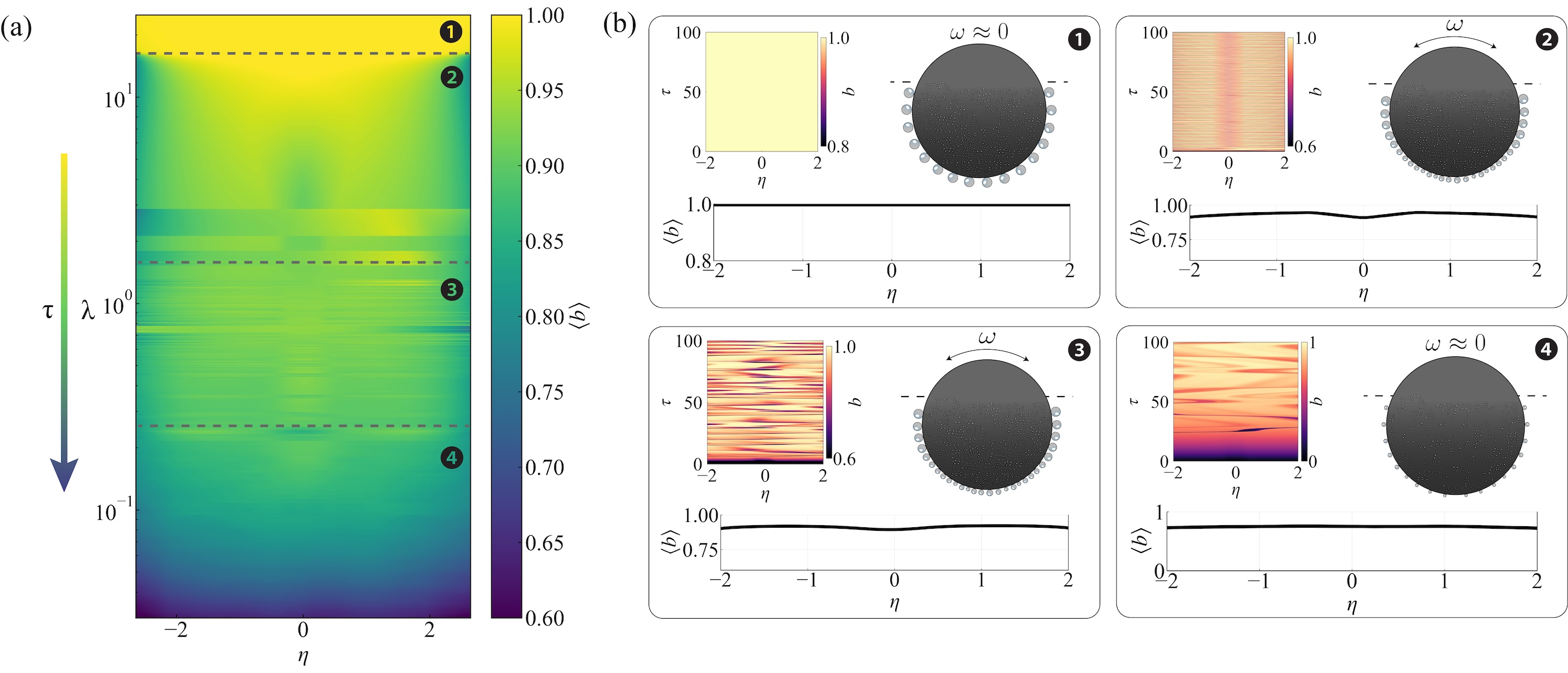}
    \caption{Asymmetric bodies. (a,b) As in Fig.~\ref{fig: Figure6}, with the same parameters, but with $\delta=100$. The dynamics in stages 1 and 3 remain qualitatively similar to those for symmetric bodies. In stage 2, however, the gravitational restoring torque due to the mass asymmetry precludes a steady rolling state. Instead, the system settles into a limit cycle, with either periodic back-and-forth oscillation, or unidirectional rolling with undulations; both types can be seen in (a). In stage 4, capsizing events are restricted by bottom-heaviness, replaced by a gentle swinging motion.}
    \label{fig: Figure7}
\end{figure*}

At late times, or with very low growth rates, the dynamics enter a quieter fourth stage: the cylinder is mostly stationary, interrupted by isolated rolling (capsizing) events. After a long, slow growth period, there is eventually sufficient bubble coverage to destabilize the cylinder, and one large, bubble-clearing event commences. Each rolling event carries the air-exposed patch down into the fluid, which comes to rest at similar locations after each such rotation, resulting in a diminished mean surface coverage near the lowest point on the body.

In all four stages, when the cylinder rotates quickly, some bubbles can survive their passage out of the water and back into the bath before bursting, so that no part of the body is ever left completely free of bubbles (as observed in the experiments; see Movie M4). In stage 1 this plays no role, since the rapid growth keeps the surface uniformly covered in any case. In the other stages, however, the effect of these surviving bubbles depends on whether the rotation is steady. In stage 2 the rotation rate is constant, so the surviving bubbles act as a fixed bias on the field and do not disturb the steady rolling. In stages 3 and 4 the rotation rate varies and reverses, so this bias is no longer constant, and it adds to the irregularity of the bubble field and reinforces the chaotic dynamics.

A common behavior seen in the experiments but not accounted for above, however, is body oscillation, both periodic and aperiodic. Such behavior, then, appears to be due to a non-zero center of mass offset, a possibility to which we now turn.

\subsection{Extreme sensitivity to body mass distribution}\label{sec:Robbins}
We now consider a non-zero center of mass offset, $\delta\neq 0$. Figure \ref{fig: Figure7}(a,b) shows how a center-of-mass offset reshapes the bubble-field dynamics. As $\lambda$ decreases, altered versions of the four life stages identified for symmetric bodies persist. In the first stage, the behavior remains unremarkable. The bubble growth rate is large, the cylinder is stable to any rotational perturbations, and the surface remains uniformly saturated at equilibrium. Here, however, the body orientation is selected by transient gravitational relaxation towards $\theta=0$ (where the center of mass rests at its lowest possible point), rather than by initial conditions. 

The most noticeable alteration to the dynamics is observed in the second stage. As discussed in \S\ref{sec:Reduced order}, at $\R=1+\chi \tilde{\delta}/(\Pr+1)$ the system undergoes a bifurcation \cite{Sparrow1982}. But the steady rolling states for symmetric bodies are now impossible; the bifurcation is now a Hopf bifurcation, and the steady rolling solution is replaced by a limit cycle. The limit cycle corresponds either to unidirectional rolling, with periodic surges in rotation rate (as the effective pendulum swings through its lowest point), or, if the buoyancy is not strong enough to overcome the mass asymmetry, to periodic rocking dynamics. Both behaviors are visible in the pointwise, time-averaged bubble coverage in Fig.~\ref{fig: Figure7}a. The corresponding bubble field, shown in Fig.~\ref{fig: Figure7}b (a rocking case), shows coherent, oscillatory banding of $b(\eta, \tau)$ in time. In addition, since the rotation reverses periodically rather than persisting in one direction, $\langle b \rangle(\eta)$ is approximately symmetric about $\eta=0$ in this case. 

In the third stage, the cylinder rotates chaotically, reversing direction with no preferred direction, as in the case for symmetric bodies. During this stage, the trajectory passes through the parameter window that Sparrow~\cite{Sparrow1982} associates with a homoclinic explosion, in which a dense family of unstable periodic orbits is created and the transient dynamics become increasingly erratic~\cite{Kaplan1979,Strogatz2015}. Since the body is almost always rotating (in one direction or another, and at widely varying rates), the restoring torque from the center of mass is nearly always engaged, acting as an additional unsteady forcing on top of an already chaotic system. The bubble field is correspondingly more complex and less symmetric when compared to the corresponding fields in Fig.~\ref{fig: Figure6}b. 

Finally, in the fourth stage, any bottom-heaviness can encourage calmer behavior. Rather than the large, sudden capsizing events in the late stage for symmetric bodies, the gravitational torque resists such rotational events, resulting in a slower swinging, which gently removes the bubbles residing closest to the waterline. The time-averaged coverage is therefore more saturated than in the case of symmetric bodies, and the central depletion seen in the last panel of Fig.~\ref{fig: Figure6}b has been removed.

Taken together, the four stages show that even a very small mass asymmetry can have a profound effect on the dynamics, rather than just perturbing the behaviors observed for symmetric bodies. Related systems (which do not include gaseous fluids) have also shown an extreme sensitivity to mass asymmetry \cite{wk21,awnlvk24,xie2024subtle}. Hence, the small gravitational contribution from imperfections in a real-world manifestation of this system should be included in any modeling effort.

\section{Discussion}

We have shown by experiment and theory that a floating body atop a supersaturated fluid displays a variety of rich dynamics. The system glides autonomously through a sequence of bifurcations in time, through multiple distinct life stages, due to the steady escape of gas into the surrounding environment. Observations which were captured by the theory include minutes-long, steady rolling states, a transition to chaotic dynamics, and a strong dependence on any mass asymmetry or imperfections. Modeling the surface bubble field as a continuum presented opportunities for analytical insight into the dynamics, namely the connection to the Lorenz system.

The continuum limit represented in the model is only expected to be accurate down to cylinder radii considerably larger than the maximum bubble radius. Below this radius, the dynamics will be driven by the idiosyncratic behavior of individual bubbles, particularly large bubbles, as was the case in the examination of vertical dancing \cite{scg24}. Large bubbles also tend to detach and clear the surface of smaller bubbles while sliding upwards towards the surface, which has been studied more directly in Refs.~\cite{rm16,hsxmjej23}. This points to the potentially enhanced sensitivity of the system in this regime to the `Fritz radius' beyond which bubbles detach, which depends on surface properties and curvature \cite{Fritz35,op93}. 

Several factors were held nearly fixed in our exploration, though they are likely deserving of their own study. The body density selects how deeply the cylinder sits in the fluid. The important parameters in the Lorenz system, $\R$ and $\Pr$, and the various critical growth rates described in \S\ref{sec:Reduced order}, all depend on $\eta_s$. How the body sits in the fluid can thus strongly affect the dynamics, even if the body remains positively buoyant. Lighter large cylinders, which were roughly half-immersed at equilibrium, simply come to a rest, with only small rotational fluctuations, indefinitely. Heavier bodies, meanwhile, reintroduce the vertical `dancing' dynamics studied in Ref.~\cite{scg24}, which in tandem with the rolling behaviors presented herein, can exhibit even more exotic dynamics. 

Body geometry can also introduce new dimensions; preliminary studies with spheres has shown remarkable dynamics, with unidirectional rolling resulting in the removal of a continuous band of surface bubbles, which then introduces an instability to rotation about an orthogonal axis. Here too, a far richer set of dynamics is on offer. Surface physics, which can strongly affect the nature of bubble growth and removal, is another intriguing direction for future investigation; non-uniform or even time-dependent surface properties may allow for unique methods of controlling the dynamics. 

Also observed experimentally, but outside of the scope of this paper, was slow cylinder translation along the surface. This motion, which appeared correlated with body rotation, was reminiscent of a steamboat's wheeled propulsion. For constant rolling the body slowly moved along the fluid-air interface in one direction. We observed one case in which the cylinder reached a bounding wall, which promoted a change in the sign of the rotation rate; the body then began its journey back across the surface in the opposite direction. These dynamics appear to have some connection to the oscillatory dynamics of floating surfaces atop convection cells due to thermal blanketing, a model of continental drift \cite{zl00,zz05,Huang25}. 

Particularly during the late life stages of the bubblewheel experiment, the body behavior is reminiscent of a more dramatic occurrence at a far larger scale, the capsizing of icebergs \cite{Burton12,bmw13,byqlmcssm20}. As has been recently shown, the submerged volume of floating ice can develop sharp features as it melts, resulting in sudden body reorientation~\cite{jzkwr24,jwzkr25,nu26}. In both settings, pedestrian features above the waterline belie rich, dynamic, and potent physics, roiling just below the surface.

\section{Data availability statement}
Data from the experiment or simulations provided in the figures will be made available upon request.

\section{Acknowledgements}
The authors are grateful for the support of the National Science Foundation (NSF grant DMS-2527011), the Office of the Vice Chancellor for Research and Graduate Education with funding from the Wisconsin Alumni Research Foundation, and by donations to the Applied Math, Engineering, and Physics (AMEP) program at the University of Wisconsin–Madison. The authors also acknowledge helpful conversations with John W.~M.~Bush and Jean-Luc Thiffeault.

\section{Author contribution statement}
M.~Z. and S.~E.~S. contributed equally to this work. M.~Z. constructed and carried out the experiments. M.~Z. and S.~E.~S. conceptualized the work, wrote the code, and wrote the paper.

\bibliographystyle{unsrt}
\bibliography{Bib}

\newpage
\begin{widetext}

\setcounter{figure}{0}
\renewcommand{\thefigure}{S\arabic{figure}}
\setcounter{equation}{0}
\renewcommand{\theequation}{S\arabic{equation}}
\setcounter{table}{0}
\renewcommand{\thetable}{S\arabic{table}}

\begin{center}
    \textbf{Autonomous route to the strange attractor: \\the many life stages of the chaotic bubblewheel - \\ Supplementary material}%
\end{center} 

\section{Table of parameters}

The four bodies used in experiments are as follows. Body I is a large, symmetric cylinder. Body II is a large, slightly bottom-heavy cylinder. Body III is a large, more substantially bottom-heavy cylinder. Body IV is a small, symmetric cylinder. Their geometric and measured parameters are given in Table~\ref{tab:parameters}, along with fitted and derived parameters, computed as described below. Since the initial gas concentration can vary substantially, we give parameters associated with low and high values of the initial bubble growth rate of $\Lambda = 4$ dyn cm$^{-2}$ s$^{-1}$ and $\Lambda = 50$ dyn cm$^{-2}$ s$^{-1}$, respectively.

\begin{table}[h]
    \caption{Geometric and inertial body properties.}
    \label{tab:parameters}
    \centering
    \renewcommand{\arraystretch}{1.2}
    \begin{tabular}{lcccc}
        \toprule
        & \multicolumn{4}{c}{Body}\\
        \cmidrule(lr){2-5}
        & I & II & III & IV\\
        \midrule
        \multicolumn{5}{l}{\emph{Measured}}\\
        Radius, $a$ [cm]                            & 2.5 & 2.5 & 2.5 & 1.5\\
        Length, $L$ [cm]                            & 8.5 & 7.55 & 8.5 & 8.5\\
        Mass, $m$ [g]                               & 134.7 & 139.9 & 143.0 & 50.1\\
        Effective density, $\rho_s$ [g\,cm$^{-3}$]  & 0.81 & 0.94 & 0.86 & 0.83\\
        Timescale, $t_c$ [s]                    & 63.1 & 73.7 & 66.9 & 23.5\\
        Bubble forcing, $\beta$                    & $1.86\times10^{4}$ & $2.30\times10^{4}$ & $2.02\times10^{4}$ & $7.00\times10^{3}$\\
        \midrule
        \multicolumn{5}{l}{\emph{Fitted}}\\
        Center-of-mass offset, $d/a$   & $0.05\%$ & $0.5\%$ & $6\%$ & $0.08\%$\\
        Scaled offset, $\delta$        & $1.0\times10^{3}$ & $1.26\times10^{4}$ & $1.33\times10^{5}$ & $3.7\times10^{2}$\\
        Drag coefficient, $\Sigma$     & 27.53 & 23.41 & 25.34 & 23.51\\
        Scaled drag coefficient, $\sigma$     & 17.0 & 15.3 & 16.0 & 14.7\\
        Removal rate, $\lambda'$       & 5.3 & 6.4 & 6.4 & 3.2\\
        \midrule
        \multicolumn{5}{l}{\emph{Derived, at $\Lambda = 4\,\mathrm{dyn\,cm^{-2}\,s^{-1}}$ (early time approximation, low end)}}\\
        Initial dimensionless bubble growth rate, $\lambda$                    & 4.2 & 4.9 & 4.5 & 1.6\\
        Prandtl number, $\mathrm{S} = \sigma/\lambda$ & 4.04 & 3.11 & 3.59 & 9.40\\
        Rayleigh number, $\mathrm{R}$           & 296 & 348 & 322 & 347\\
        Reduced offset, $\tilde\delta$          & 17.5 & 161 & $2.07\times10^{3}$ & 46.7\\
        \midrule
        \multicolumn{5}{l}{\emph{Derived, at $\Lambda = 50\,\mathrm{dyn\,cm^{-2}\,s^{-1}}$ (early time approximation, high end)}}\\
        Initial dimensionless bubble growth rate, $\lambda$                    & 52.5 & 61.4 & 55.8 & 19.5\\
        Prandtl number, $\mathrm{S} = \sigma/\lambda$ & 0.324 & 0.249 & 0.287 & 0.752\\
        Rayleigh number, $\mathrm{R}$           & 23.7 & 27.8 & 25.7 & 27.7\\
        Reduced offset, $\tilde\delta$          & 0.11 & 1.03 & 13.20 & 0.30\\
        \bottomrule
    \end{tabular}
\end{table}

\section{Body tracking}

The body orientation and rotation rate were inferred from overhead
video recordings using computer vision. Each video was recorded at $60$ frames per
second. The procedure consists of three stages: identification of the cylinder and its rotation axis in each frame, reconstruction of the trajectories of visual markers attached to the body surface, and conversion of the marker velocities into an instantaneous rotation rate. Throughout, positions are measured in pixels in the image plane, and the frame index is denoted by $k$.
\begin{figure}
    \centering
    \includegraphics[width=0.9\linewidth]{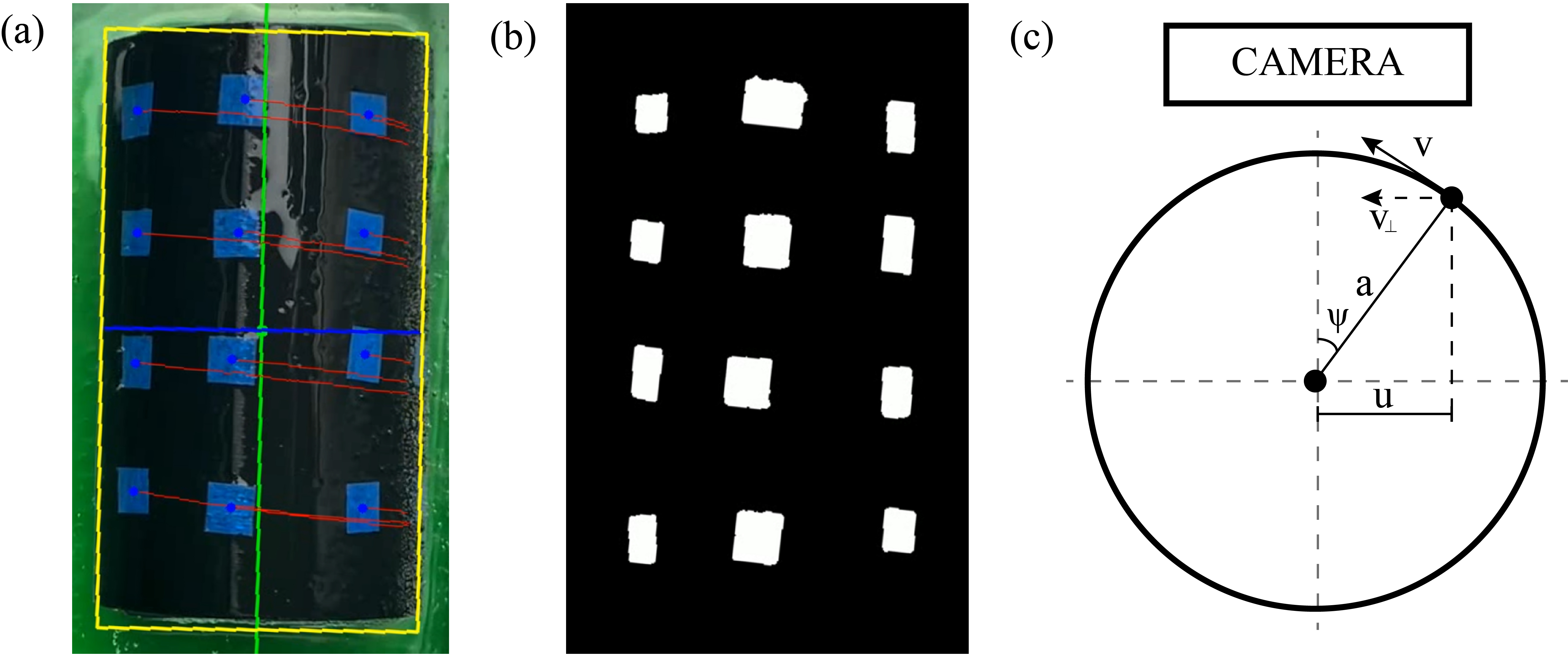}
    \caption{(a) A video frame demonstrating the computer vision pipeline for marker detection and cylinder rotation tracking. The yellow box outlines the detected cylinder region, dark blue dots indicate the centroids of identified markers, the vertical green line shows the estimated rotation axis, and red curves record the trajectories of each marker over time. (b) Binary mask generated by applying color thresholding to the original frame. White pixels are regions that pass the threshold, corresponding to the identified blue markers attached to the cylinder surface. (c) Geometry of the angular velocity calculation, viewed along the rotation axis. The circle is the cylinder cross-section of radius $a$; the dot at center is the rotation axis, and the dot on the circle is a marker. The angle $\psi$ is measured from the direction facing the camera to the marker. The marker's horizontal distance from the axis, $u = a\sin\psi$, is the quantity measured in the overhead frame, while its distance from the axis toward the camera, $\sqrt{a^2-u^2} = a\cos\psi$, is not measured. The marker's velocity $\bm v$ is tangent to the circle; its component $v_\perp$, the only part visible from above, is what enters the angular velocity calculation.}
    \label{fig: FigureS1}
\end{figure}

\textbf{Step 1: Cylinder identification.} In each frame, the cylinder was identified by exploiting its color contrast with the background. Since the cylinder was painted black and the background was bright and highly saturated, the 2D projection of the cylinder appeared as a compact rectangular region of pixels with low saturation and low value. Thresholds were thus applied on the saturation and value channels in HSV space to obtain a mask of candidate pixels. Morphological operations were then used to remove noisy pixels from bubble reflections and to fill small gaps. From the processed mask, we then extracted all connected components and identified the component with the largest rectangular contour as the cylinder. The centroid of this contour was taken as the cylinder center in the image plane, and the major axis of the contour defined the in-plane orientation of the cylinder. Fig.~\ref{fig: FigureS1}a illustrates a representative identification result.

\textbf{Step 2: Trajectory tracking of markers.} The trajectory and rotational motion of the cylinder were inferred from blue visual markers attached to its side. These markers were isolated by applying a color threshold in HSV space, followed by morphological opening and closing to suppress bubble reflections and other small artifacts. Each marker then appears as a compact connected region in the binary mask, as shown in
Fig.~\ref{fig: FigureS1}b. For each region whose area exceeded a minimum threshold, we computed its centroid and treated it as a valid marker detection in that frame.

Marker trajectories were reconstructed by establishing correspondences between marker detections across consecutive frames. Since the cylinder rotates smoothly and the frame rate is high, the displacement of a marker between consecutive frames is small compared with the spacing between neighboring markers. Thus, for each marker detected in frame $k$, we searched in frame $k+1$ for the nearest marker detection within a prescribed maximum displacement threshold. To reject spurious matches, we additionally required that the marker area did not change by more than a specified fraction. If a marker in frame $k$ did not find any detection in frame $k+1$ satisfying these criteria, we assumed that it had rotated out of view and terminated its tracking. Conversely, unmatched detections in frame $k+1$ were treated as newly visible markers, and new tracks were initialized accordingly. An example of the reconstructed marker trajectories is shown in Fig.~\ref{fig: FigureS1}b.

\textbf{Step 3: Angular velocity estimation.} For each tracked marker, instantaneous velocity was obtained by differencing its position trajectory, and only its horizontal component $v_\perp$ is measured by the overhead camera. As shown in Fig.~\ref{fig: FigureS1}c, a marker at horizontal distance $u$ from the rotation axis lies at depth $\sqrt{a^2-u^2}$ along the line of sight, and since the marker moves on a circle with radius $a$, this depth relates to the measured velocity by $v_\perp = \Omega\sqrt{a^2-u^2}$. At each frame we therefore projected the marker velocity onto the horizontal direction, subtracted the velocity of the cylinder center, and divided by $\sqrt{a^2-u^2}$ to obtain $\Omega$, discarding markers with $|u|$ too close to $a$. The rotation rate was then taken as the median over all valid markers.

\section{Measuring the mass asymmetry and rotational drag in gas-free water}
 
The mass distribution inside the 3D printed bodies was not easily controlled with precision with the 3D printing approach taken. The center of mass offset $d$ and the drag coefficient $\Sigma$ appearing in Eq.~(2) of the main text are therefore not known in advance, and were instead determined for each body by a rolling relaxation test using gas-free tap water. In the absence of dissolved gas, no bubbles form on the submerged surface, the surface buoyancy torque $T_b$ vanishes, and the angular momentum balance (1) of the main text reduces to that of a damped pendulum,
\begin{equation}
    \frac{m a^{2}}{2}\ddot{\theta} = -mgd\sin\theta - 4\pi\mu a^{2}L\,\Sigma\,\dot{\theta} - \frac{\rho\pi a^{4}L}{4}\ddot{\theta}.
    \label{eq:relax_dim}
\end{equation}
Scaling time upon $t_c = \rho_s a^{2}/8\mu$ as in the main text, and writing $\tau = t/t_c$, this becomes
\begin{equation}
    \theta_{\tau\tau} = -\delta\sin\theta - \sigma\,\theta_{\tau},
    \label{eq:relax}
\end{equation}
with $\delta$ and $\sigma$ as defined in (8) and (9) of the main text.
 
The offset $d$ admits a simple interpretation. Consider an internal density profile that is linear across the cross-section,
\begin{equation}
    \rho_s(\bm{x}) = \rho_0 + \rho_1\, \bm{p}\cdot\frac{\bm{x}-\bm{x}_0}{a},
\end{equation}
with $\bm{p}(\theta) = (\sin\theta,-\cos\theta)$ as in the main text. Then with $m = \pi a^{2}L\rho_0$, the moment of inertia is given by $ma^2/2$, up to a correction of relative order $(d/a)^2$, and the gravitational torque is $-mgd\sin\theta$ with
\begin{equation}
    d = \frac{a}{4}\frac{\rho_1}{\rho_0}.
\end{equation}
The density becomes negative inside the body if $\rho_1 > \rho_0$, so this model requires $d/a < 1/4$. The values realized in the experiments are all smaller than this constraint, by at least a factor of four.

\subsection{Fitting procedure}

For each test, the body was placed at the center of the bath, given a small initial rotation rate by hand, and released, and was then allowed to come to rest on its own. The relaxation was recorded from above and the rotation rate was extracted by the same pipeline described above. Since no bubbles form in gas-free water, the only parameters governing the motion are the center of mass offset, $\delta$, and the viscous drag, $\sigma$. Both were obtained for each body by fitting the solution of \eqref{eq:relax} to the recorded trials.

The fit then proceeds as follows. A range of possible values was first prescribed for both parameters. For every candidate set drawn from this range, equation \eqref{eq:relax} was integrated forward in time and the resulting $\Omega(t)$ was interpolated onto the times at which the experimental data were sampled, and the sum of squared residuals between the two was evaluated. The set giving the smallest residual returns the reported values for that trial. Several trials are computed and the final fitted value for that tested body is the average of them.

\begin{figure}
    \centering
    \includegraphics[width=0.9\linewidth]{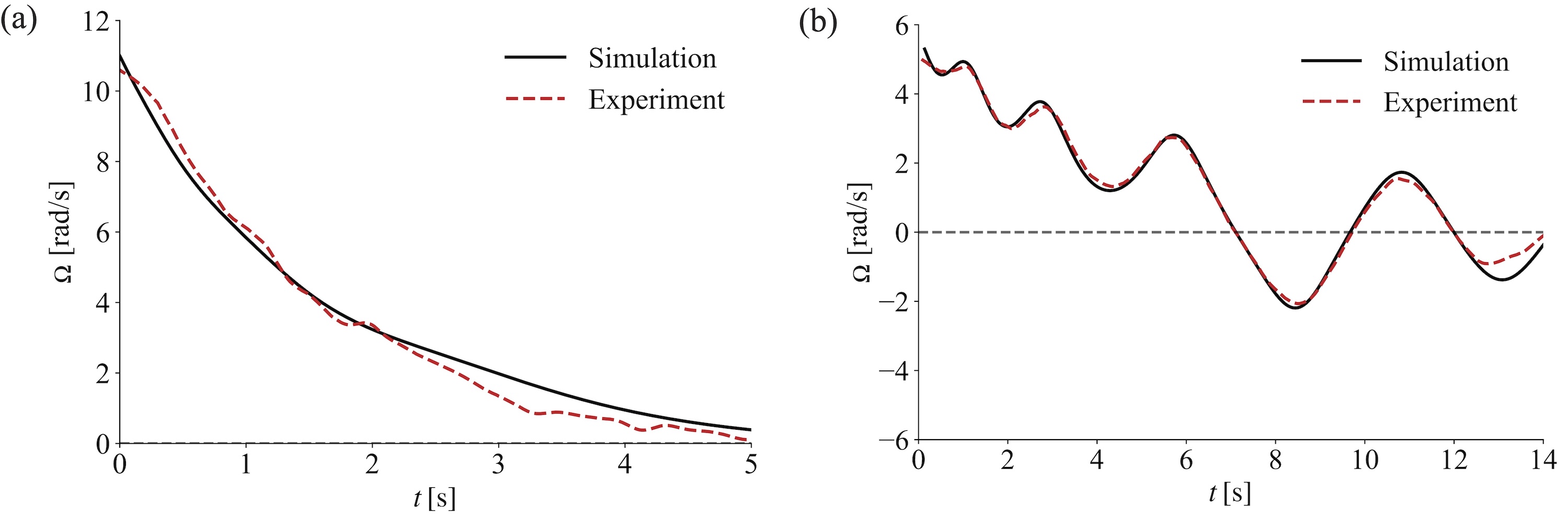}
    \caption{ Rotational relaxation on the surface of gas-free tap water, used to determine $\delta$ and $\sigma$. (a) Calibration for the small sized, symmetric Body IV; the rotation rate decays monotonically to rest over the record. (b) Calibration for the large, slightly bottom-heavy Body II, which introduces more prominent body oscillations during its relaxation.}
    \label{fig: FigureS2}
\end{figure}

Figure~\ref{fig: FigureS2} shows the outcome of this procedure for two bodies whose relaxations are qualitatively different. The rotation rate of the small, symmetric Body IV  (Bodies I - IV are defined above) decays monotonically to rest, so that the relaxation is governed almost entirely by the viscous term in \eqref{eq:relax}. The large, slightly bottom-heavy Body II instead swings back through its equilibrium orientation several times before coming to rest. In both cases, the fitted curves follow the experimental results closely. The parameters used for both fits are given in Table~\ref{tab:parameters}.

\section{Determination of the bubble removal rate}

With $\delta$ and $\sigma$ fixed by the gas-free calibration above, two unknowns remain in the full model: the bubble growth rate $\lambda$ on the submerged surface, and the bubble removal rate $\lambda'$ on the exposed surface. Both are obtained by fitting solutions of Eqs.~(6) and (7) in the main text to trials conducted in carbonated water.

Dissolved gas escapes continuously over the life of an experiment, so the (scaled) growth rate, $\lambda$, decreases over time. A fitting window taken shortly after immersion therefore has a different $\lambda$ than a window taken several minutes later, even within a single trial. Assigning one global value of $\lambda$ to the entire record would force this time dependence into the fit of $\lambda'$ instead, producing a biased removal rate that compensates for a growth-rate error rather than reflecting the true bubble dynamics.

To avoid this, $\lambda$ is fitted locally. Each trial is divided into short windows, and within each window $\lambda$ is treated as constant and fitted jointly with $\lambda'$ to reproduce the measured rotation rate over that interval. The free parameters within each window are therefore the growth rate $\lambda$, the removal rate $\lambda'$, and the initial orientation $\theta(0)$, since the body's orientation at the start of a given window is not independently measured and must be recovered from the fit itself. The offset $\delta$ and drag coefficient $\sigma$ are held fixed at the values obtained in the previous section, and residuals are minimized using the same algorithm described there for the gas-free tests.


Table~\ref{tab:parameters} includes the fitted parameters above. Derived parameters needed in the main text are also reported. For these derived parameters, we use a reference low-end, early-time bubble growth rate, $\Lambda = 4\,\mathrm{dyn\,cm^{-2}\,s^{-1}}$, used for the model curves in Fig.~2c of the main text, taking $b_s = 60\,\mathrm{dyn\,cm^{-2}}$, $\mu = 0.01\,\mathrm{P}=0.01\,\mathrm{dyn\, s/cm}^2$ and $\eta_s = 2.8$. We also include derived parameters for a high-end early time growth rate, $\Lambda = 50\,\mathrm{dyn\,cm^{-2}\,s^{-1}}$. Since $\Lambda$ decays over the course of an experiment, these are representative early-time values rather than fitted ones, and they scale as $\lambda \propto \Lambda$, $\mathrm{S} \propto \Lambda^{-1}$, $\mathrm{R} \propto \Lambda^{-1}$ and $\tilde\delta \propto \Lambda^{-2}$.

\section{Size dependence of the steady rolling rate}

Figure~\ref{fig: FigureS4} compares the rotation rate for the two nearly uniform bodies, the larger Body I ($a=2.5$cm) and the smaller Body IV ($a=1.5$cm). The mean of the rotation rate is comparable, $\langle \Omega\rangle = 0.94$ rad/s for Body I and $\langle \Omega\rangle = 1.07$ rad/s for Body IV.

\section{From steady rolling to chaos}
\begin{figure}
    \centering
    \includegraphics[width=0.5\linewidth]{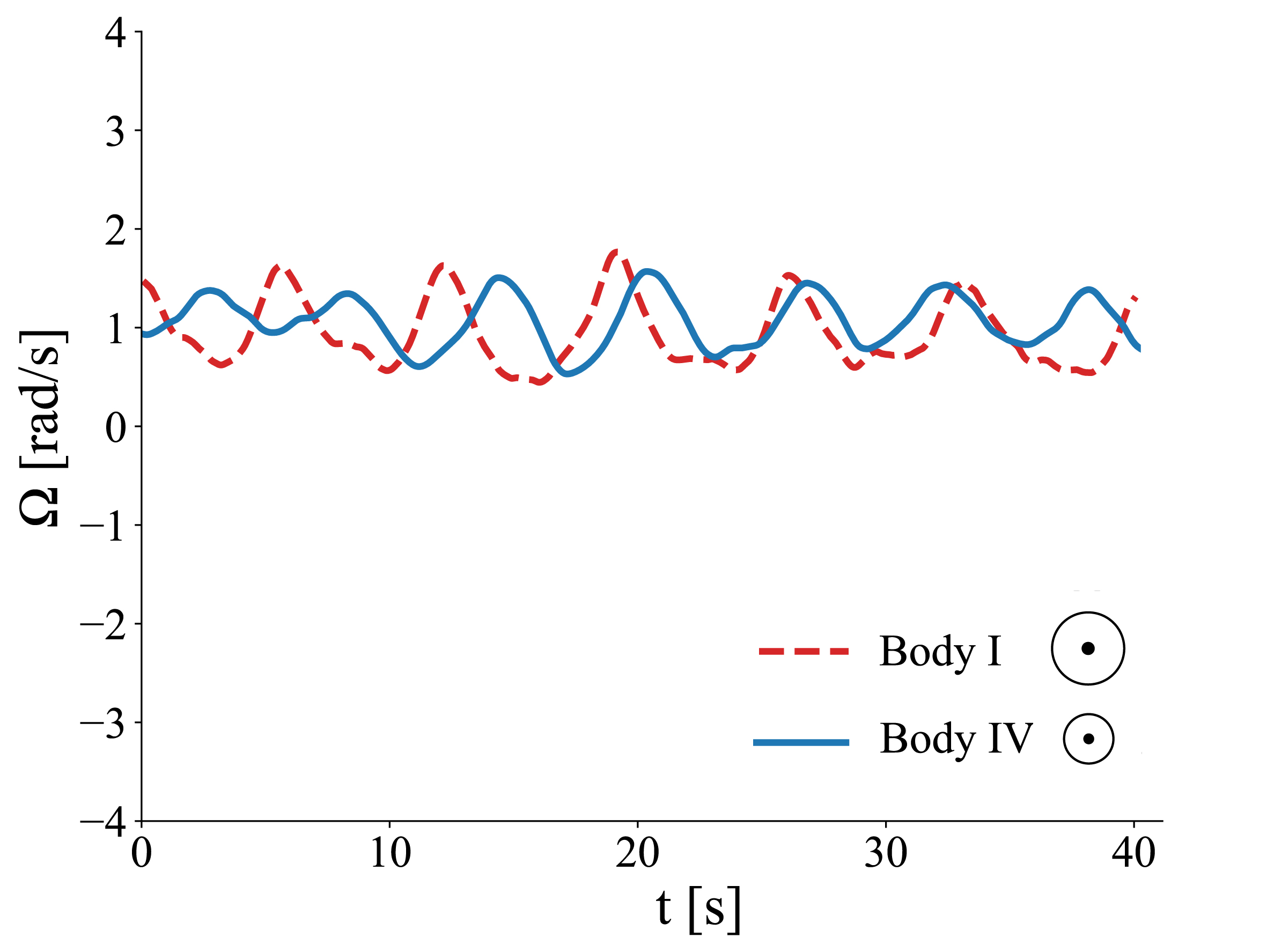}
   \caption{Rotation rate $\Omega(t)$ for the large, uniform Body I and the smaller, uniform Body IV, captured at comparable gas concentration. Despite having different radii ($a=2.5$cm and $a=1.5$cm, respectively), the two bodies exhibit similar rotation rates. The averaged rotation rate over a one minute window (from $t=1$min to $t=2$min) is 0.94 rad/s for Body I and 1.07 rad/s for Body IV. }
    \label{fig: FigureS4}
\end{figure}

\begin{figure}
    \centering
    \includegraphics[width=0.5\linewidth]{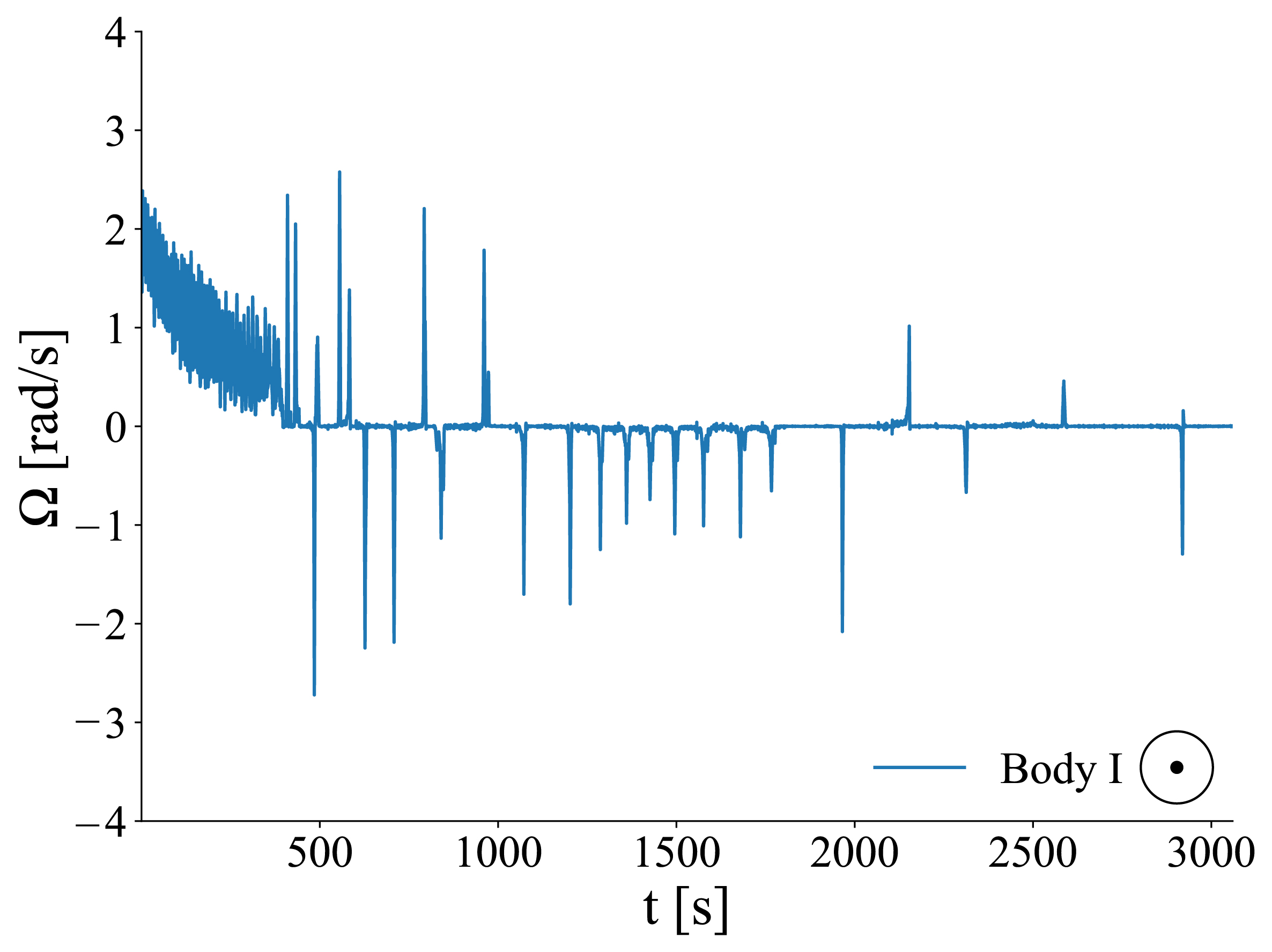}
    \caption{Rotation rate $\Omega(t)$ for a single extended trial of the large, symmetric Body I
($a=2.5$cm, $L=8.5$cm, $\rho_s=0.81$, $d/a=0.05\%$), recorded continuously for 50 minutes. For $t\lesssim400$s the rotation rate remains positive and the body rolls unidirectionally. The rotation rate first changes sign near $t=430$s. The following stage consists of isolated excursions of both signs with no obviously preferred direction. Beyond $t\approx1800$s these events become progressively rarer, since the surface requires a time $\sim b_s/\Lambda$ to reaccumulate.}
    \label{fig: FigureS3}
\end{figure}

Most of the trials reported in the main text lasted between 10 and 15 minutes. However, to capture the entire sequence of dynamical regimes within one continuous record, a single trial using Body I ($a=2.5$cm, $L=8.5$cm, $\rho_s=0.81$, $d/a=0.05\%$) was extended to just over 50 minutes. The measured rotation rate is shown in Fig.~\ref{fig: FigureS3}. 

For the first 400 seconds the rotation rate never changes sign. The body rolls continuously in the direction selected upon immersion, and the rate oscillates without reversing. This time period corresponds to the `second stage' of the dynamics, described in the main text: the steady rolling state that emerges from the pitchfork bifurcation. Later, after sufficient gas has escaped from the fluid, the rotation rate changes sign (at $t=430$s). Beyond this point the experiment consists of isolated rolling events in one direction or the other, separated by intervals of near rest, with no preferred direction and no repeating period. This is closer to the `fourth stage' behavior discussed in the main text. We hypothesize that the chaotic third stage persists for a relatively short time window in this experiment. 

\section{Critical center-of-mass offset to stabilize rotation}

An asymmetric cylinder can remain stable under perturbation by rotation when the center-of-mass offset, $d/a$, is sufficiently large. Expanding the generalized Lorenz system in the main text about the origin, we inspect the Jacobian of the linear system, whose eigenvalues are the roots of the characteristic polynomial
\begin{gather}
    p(\xi)=(1+\xi)\left(\xi ^3+ (1+\mathrm{S})\xi ^2+ (\chi\tilde{\delta}-\mathrm{R}\mathrm{S}+\mathrm{S})\xi+\chi\tilde{\delta}\right).
\end{gather}
All roots of a cubic polynomial $\xi^3+a_2 \xi^2+a_1\xi+a_0$ have negative real part if and only if $a_2>0$, $a_0>0$, and $a_1a_2>a_0$ (see \cite{Gantmacher59}); or in the present case, once $ \tilde{\delta}=(1 + \mathrm{S}) (\mathrm{R} - 1)/\chi$. So for $\mathrm{R}>1$, where unsteady rolling/rocking, or chaotic dynamics would normally appear, if $\chi \tilde{\delta}$ is greater than the value above, the body instead is predicted to remain motionless. Inserting physical parameters, this occurs at the critical center-of-mass offset of
\begin{gather}\label{dstar_full}
    \frac{d}{a} = \frac{\left(a^2 \Lambda(\rho +2 \rho_s)+16 b_s \mu  \Sigma \right)\left(3 k b_s^2-4 \pi  \eta_s  \Lambda  \mu  \Sigma \right) }{16\pi  a g \eta_s b_s^2\rho_s \mu \Sigma }.
\end{gather}
In the range of parameters relevant to the experiments at early times, where $3 k b_s^2 \ggg 4 \pi  \eta_s  \Lambda  \mu  \Sigma$, we have the cleaner approximation cited in the main text, 
\begin{gather}
\frac{d}{a} \approx \frac{3 a k}{8 \pi \mu g \sigma \eta_s }\left(\Lambda+ \frac{\sigma b_s}{t_c}\right).
\end{gather}
For a body with radius $a=2.5$cm, like Bodies I-III in the main text, using $\rho_s=0.9$g/cm$^3$ and $\eta_s=2.8$, \eqref{dstar_full} predicts a critical value of $d/a\approx 1.3\%$ at
$\Lambda = 4\,\mathrm{dyn\,cm^{-2}\,s^{-1}}$ (rising roughly to $4\%$ at
$\Lambda = 50\,\mathrm{dyn\,cm^{-2}\,s^{-1}}$), which explains the suppression of dynamics for Body III, which has $d/a\approx 6\%$.

\section{Supplementary video descriptions}

\noindent \textbf{Supplementary Movie M0:} The chaotic bubblewheel\\

\noindent Description: A large cylinder (with a slight center-of-mass offset) floats atop a supersaturated (carbonated) fluid. Bubbles which grow on the surface drive body rotation, in this case leading to chaotic rocking and rolling.\\

\noindent \textbf{Supplementary Movie M1:} Dynamical modes of a symmetric cylinder (Body I)\\

\noindent Description: A selection of dynamical behaviors are shown for a large, symmetric cylinder (Body I), extracted from a single hour-long experiment, as gas slowly escapes from the fluid into the air above.\\

\noindent \textbf{Supplementary Movie M2:} Dynamical modes of a bottom-heavy cylinder (Body II)\\

\noindent Description: A selection of dynamical behaviors are shown for a large, slightly bottom-heavy cylinder (Body II).\\

\noindent \textbf{Supplementary Movie M3:} Large mass asymmetry suppresses dynamics (Body III)\\

\noindent Description: A large center-of-mass offset (using Body III) suppresses the dynamics.\\

\noindent \textbf{Supplementary Movie M4:} Surviving the transit\\

\noindent Description: Sufficiently rapid body rotation, for a deeply seated body allows many bubbles to remain attached even as they are transported over the waterline.

\end{widetext}

\end{document}